\documentclass[10pt,conference,letterpaper]{IEEEtran}

\usepackage[acronyms,nonumberlist,nopostdot,nomain,nogroupskip]{glossaries}
\usepackage[numbers,sort&compress]{natbib}
\usepackage{xspace}
\AtBeginDocument{%
  \providecommand\BibTeX{{%
    \normalfont B\kern-0.5em{\scshape i\kern-0.25em b}\kern-0.8em\TeX}}}
\usepackage{bm}
\usepackage[utf8]{inputenc}
\usepackage[table,xcdraw]{xcolor}
\usepackage{amsmath}
\usepackage{amssymb}

\usepackage[acronyms,nonumberlist,nopostdot,nomain,nogroupskip]{glossaries}
\usepackage{booktabs}

\usepackage[acronyms,nonumberlist,nopostdot,nomain,nogroupskip]{glossaries}

\usepackage{tabularx}
\usepackage{graphicx}
\usepackage{epstopdf}

\usepackage{tikz}
\usepackage{arydshln}
\usepackage{pgfplots}
\pgfplotsset{compat=newest}
\pgfplotsset{plot coordinates/math parser=false}
\newlength\fheight
\newlength\fwidth
\usetikzlibrary{plotmarks,patterns,decorations.pathreplacing,backgrounds,calc,arrows,arrows.meta,spy,matrix}
\usepgfplotslibrary{patchplots,groupplots,colorbrewer}
\usepackage{tikzscale}
\usepackage[hidelinks]{hyperref}

\newif\ifexttikz
\exttikzfalse

\ifexttikz
	\usetikzlibrary{external}
\fi

\usepackage{multirow}

\renewcommand{\figurename}{Fig.}
\usepackage[font=footnotesize]{subcaption}
\usepackage[font=footnotesize]{caption}

\usepackage{mathtools}

\usepackage{multirow}
\usepackage{url}
\usepackage{tabularx}
\usepackage{makecell}
\usepackage{array}
\usepackage{multibib}
\newcites{methods}{References}

\newacronym{6g}{6G}{sixth generation}
\newacronym{3gpp}{3GPP}{3rd Generation Partnership Project}
\newacronym{adc}{ADC}{Analog to Digital Converter}
\newacronym{dac}{DAC}{Digital to Analog Converter}
\newacronym{5g}{5G}{5th generation}
\newacronym{aimd}{AIMD}{Additive Increase Multiplicative Decrease}
\newacronym{am}{AM}{Acknowledged Mode}
\newacronym{amc}{AMC}{Adaptive Modulation and Coding}
\newacronym{aoa}{AoA}{Angle of Arrival}
\newacronym{aod}{AoD}{Angle of Departure}
\newacronym{aqm}{AQM}{Active Queue Management}
\newacronym{awgn}{AGWN}{Additive White Gaussian Noise}
\newacronym{balia}{BALIA}{Balanced Link Adaptation}
\newacronym{bdp}{BDP}{Bandwidth-Delay Product}
\newacronym{bf}{BF}{Beamforming}
\newacronym{fpga}{FPGA}{field-programmable gate array}
\newacronym{cc}{CC}{Congestion Control}
\newacronym{cdf}{CDF}{Cumulative Distribution Function}
\newacronym{cn}{CN}{Core Network}
\newacronym{cm}{CM}{confusion matrix}
\newacronym[plural=\gls{cnn}s,firstplural=convolutional neural networks (CNNs)]{cnn}{CNN}{convolutional neural network}
\newacronym{cqi}{CQI}{Channel Quality Information}
\newacronym{cp}{CP}{Control Plane}
\newacronym{csirs}{CSI-RS}{Channel State Information - Reference Signal}
\newacronym{dc}{DC}{Dual Connectivity}
\newacronym{dce}{DCE}{Direct Code Execution}
\newacronym{dci}{DCI}{Downlink Control Information}
\newacronym{dmr}{DMR}{Deadline Miss Ratio}
\newacronym{dmrs}{DMRS}{DeModulation Reference Signal}
\newacronym{e2e}{E2E}{End-to-End}
\newacronym{ecn}{ECN}{Explicit Congestion Notification}
\newacronym{ebs}{EBS}{exhaustive beam sweep}
\newacronym{edf}{EDF}{Earliest Deadline First}
\newacronym{enb}{eNB}{evolved Node Base}
\newacronym{epc}{EPC}{Evolved Packet Core}
\newacronym{es}{ES}{Edge Server}
\newacronym{fdma}{FDMA}{Frequency Division Multiple Access}
\newacronym{fdd}{FDD}{Frequency Division Duplexing}
\newacronym[firstplural=Radio Access Technologies (RATs)]{rat}{RAT}{Radio Access Technology}
\newacronym{fs}{FS}{Fast Switching}
\newacronym{txer}{TX}{transmitter}
\newacronym{rxer}{RX}{receiver}
\newacronym{bt}{BT}{beam tracking}
\newacronym{ftp}{FTP}{File Transfer Protocol}
\newacronym{gnb}{gNB}{Next Generation Node Base}
\newacronym{bs}{BS}{Base Station}
\newacronym{harq}{HARQ}{Hybrid Automatic Repeat reQuest}
\newacronym{hetnet}{HetNet}{Heterogeneous Network}
\newacronym{hh}{HH}{Hard Handover}
\newacronym{hol}{HOL}{Head-of-Line}
\newacronym{ia}{IA}{initial access}
\newacronym{imt}{IMT}{International Mobile Telecommunication}
\newacronym{iot}{IoT}{Internet of Things}
\newacronym{los}{LOS}{Line-of-Sight}
\newacronym{lte}{LTE}{Long Term Evolution}
\newacronym{m2m}{M2M}{Machine to Machine}
\newacronym{ml}{ML}{machine learning}
\newacronym{dl}{DL}{deep learning}
\newacronym{mac}{MAC}{Medium Access Control}
\newacronym{mc}{MC}{Multi-Connectivity}
\newacronym{mcs}{MCS}{Modulation and Coding Scheme}
\newacronym{mec}{MEC}{Mobile Edge Cloud}
\newacronym{mi}{MI}{Mutual Information}
\newacronym{mimo}{MIMO}{Multiple Input, Multiple Output}
\newacronym{mmwave}{mmWave}{millimeter wave}
\newacronym{mmWave}{mmWave}{Millimeter wave}
\newacronym{mptcp}{MPTCP}{Multipath TCP}
\newacronym{mr}{MR}{Maximum Rate}
\newacronym{mss}{MSS}{Maximum Segment Size}
\newacronym{mtd}{MTD}{Machine-Type Device}
\newacronym{mtu}{MTU}{Maximum Transmission Unit}
\newacronym{nfv}{NFV}{Network Function Virtualization}
\newacronym{nlos}{NLOS}{Non-Line-of-Sight}
\newacronym{nr}{NR}{New Radio}
\newacronym{ofdm}{OFDM}{Orthogonal Frequency Division Multiplexing}
\newacronym{pdcch}{PDCCH}{Physical Downlink Control Channel}
\newacronym{pdcp}{PDCP}{Packet Data Convergence Protocol}
\newacronym{pdsch}{PDSCH}{Physical Downlink Shared Channel}
\newacronym{pdu}{PDU}{Packet Data Unit}
\newacronym{pf}{PF}{Proportional Fair}
\newacronym{pgw}{PGW}{Packet Gateway}
\newacronym{phy}{PHY}{physical layer}
\newacronym{pbch}{PBCH}{Physical Broadcast Channel}
\newacronym[plural=\gls{mme}s,firstplural=Mobility Management Entities (MMEs)]{mme}{MME}{Mobility Management Entity}
\newacronym{prb}{PRB}{Physical Resource Block}
\newacronym{pss}{PSS}{Primary Synchronization Signal}
\newacronym{pucch}{PUCCH}{Physical Uplink Control Channel}
\newacronym{pusch}{PUSCH}{Physical Uplink Shared Channel}
\newacronym{rach}{RACH}{Random Access Channel}
\newacronym{ran}{RAN}{Radio Access Network}
\newacronym{red}{RED}{Random Early Detection}
\newacronym{rf}{RF}{Radio Frequency}
\newacronym{rlc}{RLC}{Radio Link Control}
\newacronym{rlf}{RLF}{Radio Link Failure}
\newacronym{rrc}{RRC}{Radio Resource Control}
\newacronym{rrm}{RRM}{Radio Resource Management}
\newacronym{rr}{RR}{Round Robin}
\newacronym{rs}{RS}{Remote Server}
\newacronym{rsrp}{RSRP}{Reference Signal Received Power}
\newacronym{rss}{RSS}{received signal strength}
\newacronym{rtt}{RTT}{Round Trip Time}
\newacronym{rw}{RW}{Receive Window}
\newacronym{rx}{RX}{Receiver}
\newacronym{sa}{SA}{standalone}
\newacronym{sack}{SACK}{Selective Acknowledgment}
\newacronym{sap}{SAP}{Service Access Point}
\newacronym{ap}{AP}{Access Point}
\newacronym{sch}{SCH}{Secondary Cell Handover}
\newacronym{scoot}{SCOOT}{Split Cycle Offset Optimization Technique}
\newacronym{sdma}{SDMA}{Spatial Division Multiple Access}
\newacronym{sinr}{SINR}{Signal to Interference plus Noise Ratio}
\newacronym{sm}{SM}{Saturation Mode}
\newacronym{snr}{SNR}{Signal-to-Noise-Ratio}
\newacronym{son}{SON}{Self-Organizing Network}
\newacronym{ss}{SS}{Synchronization Signal}
\newacronym{ssbs}{SSBs}{synchronization signal blocks}
\newacronym{ssb}{SSB}{synchronization signal block}
\newacronym{srs}{SRS}{Sounding Reference Signal}
\newacronym{sss}{SSS}{Secondary Synchronization Signal}
\newacronym{tb}{TB}{Transport Block}
\newacronym{tcp}{TCP}{Transmission Control Protocol}
\newacronym{tdd}{TDD}{Time Division Duplexing}
\newacronym{tdma}{TDMA}{Time Division Multiple Access}
\newacronym{tfl}{TfL}{Transport for London}
\newacronym{tm}{TM}{Transparent Mode}
\newacronym{trp}{TRP}{Transmitter Receiver Pair}
\newacronym{tti}{TTI}{Transmission Time Interval}
\newacronym{ttt}{TTT}{Time-to-Trigger}
\newacronym{tx}{TX}{Transmitter}
\newacronym{ue}{UE}{User Equipment}
\newacronym{ul}{UL}{Uplink}
\newacronym{uml}{UML}{Unified Modeling Language}
\newacronym{um}{UM}{Unacknowledged Mode}
\newacronym{utc}{UTC}{Urban Traffic Control}
\newacronym{vm}{VM}{Virtual Machine}
\newacronym{rsrq}{RSRQ}{Reference Signal Received Quality}
\newacronym{rssi}{RSSI}{Received Signal Strength Indicator}
\newacronym{crs}{CRS}{Cell Reference Signal}
\newacronym{nsa}{NSA}{Non Stand Alone}
\newacronym{mrdc}{MR-DC}{Multi \gls{rat} \gls{dc}}
\newacronym{endc}{EN-DC}{E-UTRAN-\gls{nr} \gls{dc}}
\newacronym{5gc}{5GC}{5G Core}
\newacronym{si}{SI}{Study Item}
\newacronym{iab}{IAB}{Integrated Access and Backhaul}
\newacronym{wf}{WF}{Wired-first}
\newacronym{hqf}{HQF}{Highest-quality-first}
\newacronym{pa}{PA}{Position-aware}
\newacronym{mlr}{MLR}{Maximum-local-rate}
\newacronym{wbf}{WBF}{Wired Bias Function}
\newacronym{mib}{MIB}{Master Information Block}
\newacronym{sib}{SIB}{Secondary Information Block}
\newacronym{kpi}{KPI}{Key Performance Indicator}
\newacronym{ppp}{PPP}{Poisson Point Process}
\newacronym{gtp}{GTP}{GPRS Tunneling Protocol}
\newacronym{amf}{AMF}{Access and Mobility Management Function}
\newacronym{dash}{DASH}{Dynamic Adaptive Streaming over HTTP}
\newacronym{http}{HTTP}{HyperText Transfer Protocol}
\newacronym{qos}{QoS}{Quality of Service}
\newacronym{udp}{UDP}{User Datagram Protocol}
\newacronym{cu}{CU}{Central Unit}
\newacronym{du}{DU}{Distributed Unit}
\newacronym{mt}{MT}{Mobile Termination}
\newacronym{sdap}{SDAP}{Service Data Adaptation Protocol}
\newacronym{tdm}{TDM}{Time Division Multiplexing}
\newacronym{fdm}{FDM}{Frequency Division Multiplexing}
\newacronym{sdm}{SDM}{Space Division Multiplexing}
\newacronym{dag}{DAG}{Directed Acyclic Graph}
\newacronym{st}{ST}{Spanning Tree}
\newacronym{ummimo}{UM-MIMO}{Ultra-massive Multiple Input, Multiple Output}
\newacronym{uavs}{UAVs}{Unmanned Aerial Vehicles}
\newacronym{wlan}{WLAN}{Wireless LAN}
\newacronym{rlnc}{RLNC}{Random Linear Network Coding}
\newacronym{drx}{DRX}{Discontinuous Reception}
\newacronym{cpu}{CPU}{Central Processing Unit}
\newacronym{txb}{TXB}{transmitter's beam}
\newacronym{rxb}{RXB}{receiver's beam}
\newacronym{sifs}{SIFS}{Short Interframe Space}
\newacronym{difs}{DIFS}{DCF Interframe Space}

\newacronym{rfid}{RFID}{Radio Frequency Identification}
\newacronym{rfp}{RFP}{radio fingerprinting}
\newacronym{sdr}{SDR}{software-defined radio}
\newacronym{dnn}{DNN}{deep neural network}

\newacronym{od}{OD}{object detection}

\newacronym{ot}{OT}{object tracking}

\newacronym{har}{HAR}{human activity recognition}
\newacronym{csi}{CSI}{channel state information}
\newacronym{cfr}{CFR}{channel frequency response}
\newacronym{fsl}{FSL}{few-shot learning}
\newacronym{pwr}{PWR}{passive Wi-Fi radar}
\newacronym{matnet}{MatNet}{matching network}
\newacronym{prnet}{ProtoNet}{prototypical network}
\newacronym{lstm}{LSTM}{long short-term memory}
\newacronym{gan}{GAN}{generative adversarial network}
\newacronym{svd}{SVD}{singular value decomposition}
\newacronym{nic}{NIC}{network interface card}
\newacronym{sgd}{SGD}{stochastic gradient descend }
\newacronym{vht}{VHT}{very high throughput }

\tikzstyle{startstop} = [rectangle, rounded corners, minimum width=2cm, minimum height=0.5cm,text centered, draw=black]
\tikzstyle{io} = [trapezium, trapezium left angle=70, trapezium right angle=110, minimum width=3cm, minimum height=1cm, text centered, draw=black]
\tikzstyle{process} = [rectangle, minimum width=2cm, minimum height=0.5cm, text centered, draw=black, alignb=center]
\tikzstyle{decision} = [ellipse, minimum width=2cm, minimum height=1cm, text centered, draw=black]
\tikzstyle{arrow} = [thick,<->,>=stealth]
\tikzstyle{line} = [thick,>=stealth]
\tikzstyle{darrow} = [thick,<->,>=stealth,dashed]
\tikzstyle{sarrow} = [thick,->,>=stealth]
\tikzstyle{larrow} = [line width=0.1mm,dashdotted,->,>=stealth]

\makeatletter
\def\grd@save@target#1{%
  \def\grd@target{#1}}
\def\grd@save@start#1{%
  \def\grd@start{#1}}
\tikzset{
  grid with coordinates/.style={
    to path={%
      \pgfextra{%
        \edef\grd@@target{(\tikztotarget)}%
        \tikz@scan@one@point\grd@save@target\grd@@target\relax
        \edef\grd@@start{(\tikztostart)}%
        \tikz@scan@one@point\grd@save@start\grd@@start\relax
        \draw[minor help lines] (\tikztostart) grid (\tikztotarget);
        \draw[major help lines] (\tikztostart) grid (\tikztotarget);
        \grd@start
        \pgfmathsetmacro{\grd@xa}{\the\pgf@x/1cm}
        \pgfmathsetmacro{\grd@ya}{\the\pgf@y/1cm}
        \grd@target
        \pgfmathsetmacro{\grd@xb}{\the\pgf@x/1cm}
        \pgfmathsetmacro{\grd@yb}{\the\pgf@y/1cm}
        \pgfmathsetmacro{\grd@xc}{\grd@xa + \pgfkeysvalueof{/tikz/grid with coordinates/major step x}}
        \pgfmathsetmacro{\grd@yc}{\grd@ya + \pgfkeysvalueof{/tikz/grid with coordinates/major step y}}
        \foreach \x in {\grd@xa,\grd@xc,...,\grd@xb}
        \node[anchor=north] at (\x,\grd@ya) {\pgfmathprintnumber{\x}};
        \foreach \y in {\grd@ya,\grd@yc,...,\grd@yb}
        \node[anchor=east] at (\grd@xa,\y) {\pgfmathprintnumber{\y}};
      }
    }
  },
  minor help lines/.style={
    help lines,
    gray,
    line cap =round,
    xstep=\pgfkeysvalueof{/tikz/grid with coordinates/minor step x},
    ystep=\pgfkeysvalueof{/tikz/grid with coordinates/minor step y}
  },
  major help lines/.style={
    help lines,
    line cap =round,
    line width=\pgfkeysvalueof{/tikz/grid with coordinates/major line width},
    xstep=\pgfkeysvalueof{/tikz/grid with coordinates/major step x},
    ystep=\pgfkeysvalueof{/tikz/grid with coordinates/major step y}
  },
  grid with coordinates/.cd,
  minor step x/.initial=.5,
  minor step y/.initial=.2,
  major step x/.initial=1,
  major step y/.initial=1,
  major line width/.initial=1pt,
}
\makeatother

\begin{document}

\newcommand{\FW}{\texttt{ReWiS}\xspace}

\title{\FW: Reliable Wi-Fi Sensing Through Few-Shot Multi-Antenna Multi-Receiver CSI Learning}

\author{\IEEEauthorblockN{Niloofar Bahadori$^{\dagger}$, Jonathan Ashdown$^{\ddagger}$ and Francesco Restuccia$^{\dagger}$\vspace{-0.3cm}}\\
\IEEEauthorblockA{
$\dagger$ Institute for the Wireless Internet of Things, Northeastern University, United States\\
$\ddagger$ Air Force Research Laboratory, United States}
Email: \{n.bahadori, frestuc\}@northeastern.edu, jonathan.ashdown@us.af.mil\vspace{-0.3cm}
}
\IEEEoverridecommandlockouts
\IEEEpubid{\makebox[\columnwidth]{Approved for Public Release; Distribution Unlimited: AFRL-2021-3408 \hfill}
\hspace{\columnsep}\makebox[\columnwidth]{ }}
\maketitle
\IEEEpubidadjcol

\begin{abstract}
Thanks to the ubiquitousness of Wi-Fi access points and devices, Wi-Fi sensing enables transformative applications in remote health care, home/office security, and surveillance, just to name a few. Existing work has explored the usage of machine learning on \gls{csi} computed from Wi-Fi packets to classify events of interest. However, most of these algorithms require a significant amount of data collection, as well as extensive computational power for additional \gls{csi} feature extraction. Moreover, the majority of these models suffer from poor accuracy when tested in a new/untrained environment. In this paper, we propose \FW, a novel framework for robust and environment-independent Wi-Fi sensing. The key innovation of \FW is to leverage \gls{fsl} as the inference engine, which (i) reduces the need for extensive data collection and application-specific feature extraction; (ii) can rapidly generalize to new environments by leveraging only a few new samples. Moreover, \FW leverages multi-antenna, multi-receiver diversity, as well as fine-grained frequency resolution, to improve the overall robustness of the algorithms. Finally, we propose a technique based on singular value decomposition (SVD) to make the \gls{fsl} input constant irrespective of the number of receive antennas.  We prototype the \FW using off-the-shelf Wi-Fi equipment and showcase its performance by considering a compelling use case of human activity recognition. Thus, we perform an extensive data collection campaign in three different propagation environments with two human subjects. We evaluate the impact of each diversity component on the performance and compare \FW with an existing convolutional neural network (CNN)-based approach. Experimental results show that \FW improves the performance by about 40\% with respect to existing single-antenna low-resolution approaches. Moreover, when compared to a CNN-based approach, \FW shows a 35\% more accuracy and less than 10\% drop in accuracy when tested in different environments, while the CNN drops by more than 45\%. To allow reproducibility of our results and to address the current dearth of Wi-Fi sensing datasets, we pledge to release our 60 GB dataset and the entire code repository to the community.
\end{abstract}

\glsresetall

\section{Introduction}

Wi-Fi has become one of the most pervasive wireless technologies ever invented. Indeed, today Wi-Fi is ubiquitous and provides wireless connectivity to almost any device of common use, including smartphones, tablets, laptop computers, and wearable devices. Just to give an idea of how fast Wi-Fi is growing, Cisco forecast that Wi-Fi 6 hotspots are expected to grow 13 fold from 2020 to 2023 \cite{cisco2020cisco}. Given their ever-increasing ubiquitousness, significant research efforts have investigated the usage of Wi-Fi waveforms to perform device-free classification, also called \textit{Wi-Fi sensing}. An excellent survey on the topic can be found in \cite{ma2019wifi}. In a nutshell, Wi-Fi sensing is based on passive monitoring of the changes in the \gls{cfr} produced by the presence of scatterers located between the Wi-Fi transmitters and the Wi-Fi receivers. These sudden changes in \gls{cfr} can be evaluated by estimating the \gls{csi} through the pilots contained in every Wi-Fi frame preamble \cite{bejarano2013ieee}. This way, highly-innovative applications such as human activity recognition, remote health monitoring, and surveillance can be implemented \cite{jiang2018smart}. Attesting to the relevance of these applications, in September 2020, IEEE 802.11 has approved a new technical group (TG) called IEEE 802.11bf \cite{TGbfPAR}. According to the website, TGbf will define modifications to state-of-the-art IEEE 802.11 standards at both the \gls{mac} and \gls{phy} to accommodate sensing operations between 1 GHz and 7.125 GHz, as well as above 45 GHz (i.e., millimeter-wave frequencies).

\begin{figure}[!ht]
    \centering
    \includegraphics[width=.95\columnwidth]{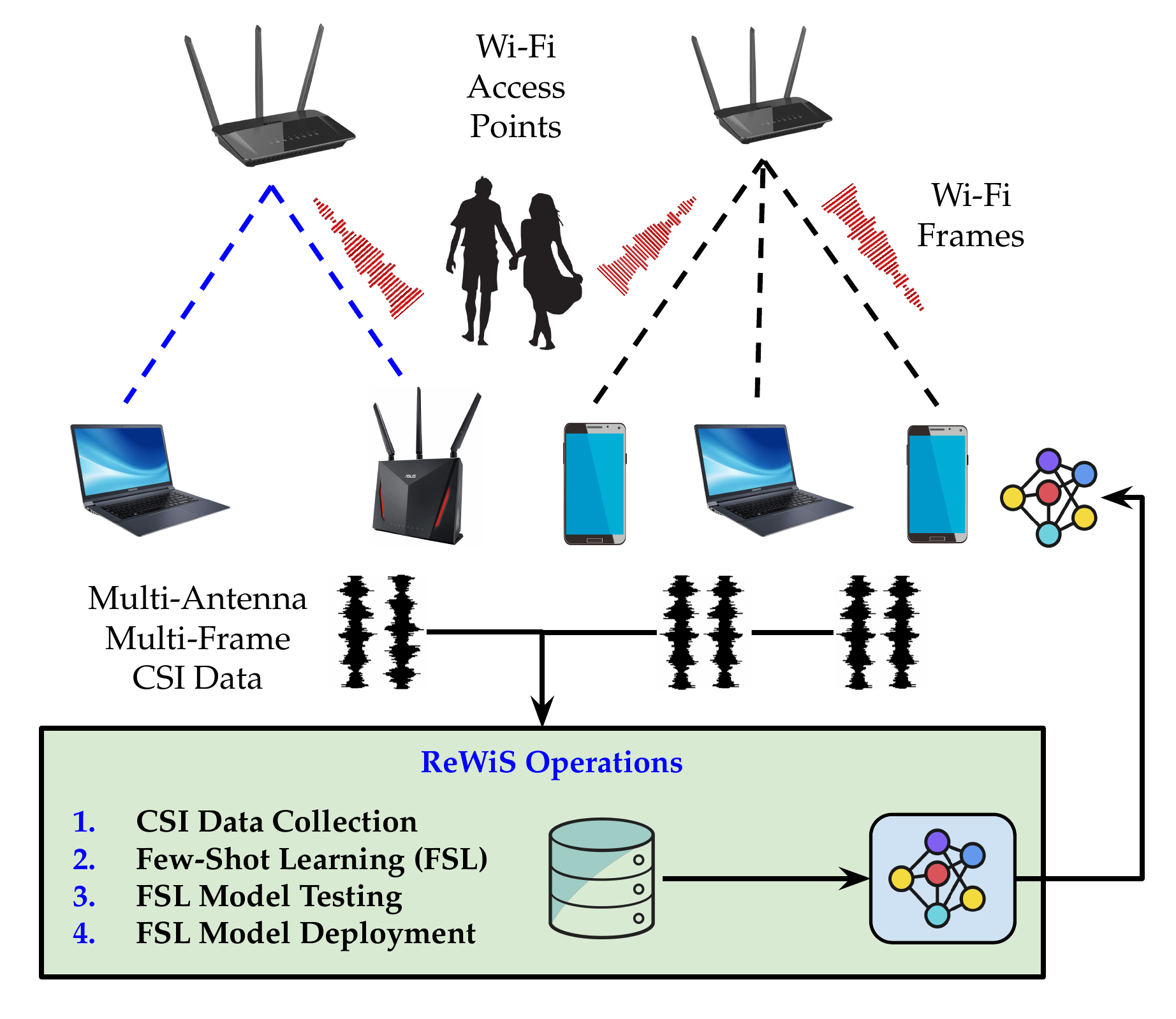}
    \caption{High-level overview of \FW.}
    \label{fig:intro}
\end{figure}

As critical Wi-Fi sensing applications come into the market, it becomes imperative to improve the \textit{robustness} of Wi-Fi sensing operations to noise and interference, as well as their amenability to \emph{generalize} to multiple operational environments. For example, if a sensing application fails to detect intruders in a smart home scenario, or stops working when deployed in a new home, it may have severe repercussions in both commercialization efforts as well as the well-being of the final users. Although existing work -- discussed in detail in Section \ref{sec:RelatedWorks} -- has proposed \gls{csi}-based sensing, they fail to generalize to multiple environments and do not address how to improve the robustness of the algorithms. 

To address the existing research gap, in this paper we propose \FW, a novel framework leveraging multi-antenna, multi-frame, multi-receiver \gls{csi} data to improve the robustness of Wi-Fi sensing operations. Figure \ref{fig:intro} shows a high-level overview of \FW and its key operations. The fundamental difference of \FW with respect to existing work is that instead of relying on traditional \gls{cnn}-based learning \cite{muaaz2021wi,BLSTM}, \FW tackles the key problem of generalization through an approach based on \gls{fsl}, which (i) reduces the need of extensive data collection; (ii) allows \FW to rapidly generalize to new tasks by only leveraging a few new samples. Moreover, \FW leverages spatial diversity (i.e., multiple receivers and multiple antennas per receiver), time diversity (i.e., multiple CSI measurements), and increased subcarrier resolution to significantly improve the robustness of the sensing process. Although existing work has proposed \gls{fsl} to address \gls{csi} learning \cite{harOneshot}, it is based on application-dependent embedding extraction using an \gls{lstm}-based technique called matching network (MatNet) \cite{matchingNet}, which ultimately limits its applicability to a single application and adds to the computational complexity of the approach. In this work, we rely on a new concept called prototypical networks (ProtoNets) \cite{snell2017prototypical} to do away with application-specific feature extraction, thus improving generalizability and computational burden.\smallskip

This paper makes the following novel contributions:\smallskip

$\bullet$ We propose \FW, a novel framework for robust and environment-independent CSI-based Wi-Fi sensing (Section \ref{sec:ProposedScheme}). The core design principles behind the \FW are to (i) leverage multi-antenna, multi-receiver diversity, as well as fine-grained frequency resolution to improve the overall robustness of the algorithms; (ii) leverage a customized version of \gls{fsl} to (a) remove the need of application-specific feature extraction; (b) help generalize to new environments by only leveraging a limited number of new samples (Section \ref{sec:rob_princ}). We also propose a technique based on \gls{svd} to make the \gls{fsl} input constant irrespective of the number of receive antennas and window size (Section \ref{sec:csi_fusion}). To give a perspective, in our dataset, we reduce the input size by about 80\% of the original size;  \smallskip

$\bullet$ We prototype \FW using off-the-shelf Wi-Fi equipment and showcase its generalizability to new environment by considering human activity recognition as a use case (Section \ref{sec:exp_data_coll}). We perform an extensive data collection campaign in three different propagation environments with two human subjects (IRB approval available upon request). We evaluate the impact of each diversity component on the performance and compare \FW with a \gls{cnn}-based approach \cite{muaaz2021wi}. Experimental results (Section \ref{sec:perf_eva}) show that the \FW improves the performance by about 40\% with respect to existing single-antenna low-resolution approaches. Moreover, when compared to a CNN-based approach, \FW shows 35\% more accuracy and less than 10\% drop in accuracy when tested in unseen environments, while the CNN performance drops by more than 45\%. \textit{To allow full reproducibility of our results, the \FW and dataset can be found at 
 }\href{https://niloobahadori.github.io/#:~:text=High\%20resolution\%20multi\%2Dreceiver\%2C\%20multi\%2Dantenna\%20CSI\%20dataset}{https://github.com/niloobah/ReWiS} \vspace{-0.1cm}

\glsresetall

\section{Background and Related Work}\label{sec:RelatedWorks}

In this section, we present some background notions on Wi-Fi sensing, as well as presenting related work and highlighting the novelty of this paper. First, we summarize Wi-Fi sensing and \gls{csi} learning in Sections \ref{sec:wifi_sensing} and \ref{sec:csi_learning}. Then, we summarize \gls{csi} collection methodologies and current \gls{csi} dataset availability in Section \ref{sec:csi_collection}.

\subsection{Wi-Fi Sensing}\label{sec:wifi_sensing}

The main idea behind Wi-Fi sensing is that the presence of moving objects alters the propagation environment throughout time and frequency. Thus, information on the presence/motion of objects in the environment can be captured by analyzing the received Wi-Fi signal. Thus, Wi-Fi sensing has been utilized in a wide range of sensing applications such as smart homes \cite{smartHomeSurvey}, device-free surveillance \cite{wifiID}, activity recognition \cite{rssiSensing,pwr, BLSTM}, and healthcare \cite{wifiRespiratoryWide,wifiMedicalResp} applications. For a comprehensive survey on the topic, please refer to \cite{ma2019wifi}. Wi-Fi sensing approaches can be categorized into three main groups, (i) \gls{rss}, (ii) \gls{pwr} and (iii) \gls{csi} sensing.  \gls{rss} sensing is performed through measuring \gls{rss} per packet, thus, is a coarse-grained parameter and provides a low-resolution information \cite{rssiSensing}. Although \gls{pwr} sensing relies on calculating the difference between transmit and receive signal, its accuracy is highly dependant on the location of the bistatic \cite{pwr}. Conversely,  thanks to the new methodologies, \gls{csi} can be captured in real-time over up to 8 antennas and up to 160MHz channels for each packet \cite{nexmonCSI}, which provides us with fine-grained information in both time and frequency domains.\vspace{-0.1cm}

\subsection{CSI Learning}\label{sec:csi_learning}

Deep learning has been proved to be an effective tool for accurate Wi-Fi \gls{csi} sensing. To improve the performance of Wi-Fi sensing in terms of accuracy, some approaches have focused on preprocessing and feature extraction \cite{doppler, dopplerSpatial, dopplerInfo, wifiRespiratoryWide,wifiMedicalResp}, while others applied more sophisticated \gls{ml} models \cite{muaaz2021wi, HarRL,BLSTM,ganHAR,harOneshot}.
However, most of these approaches are unable to generalize and their performance deteriorates when tested in an unseen environment \cite{generalizeAppDependent,BLSTM}. Commercial \gls{csi} sensing algorithms are trained offline and the source model is expected to be adapted to target environments rapidly given a limited number of samples. To this end, \textit{generalizing} the model to different environments is a crucial factor for the success of Wi-Fi sensing. A very limited number of existing papers have focused on this aspect of \gls{csi} sensing \cite{BLSTM,generalizeAppDependent,harOneshot,ganHAR}. 

Authors in \cite{BLSTM} generalize to a broader range of features by designing a very sophisticated deep learning algorithm using attention-based bi-directional long short-term memory (ABLSTM) to extract significant sequential features from raw \gls{csi} measurement. To generalize to new users, CsiGAN in \cite{ganHAR} leverages \gls{gan} to generate artificial examples to account for the left-out user whose \gls{csi} data is not available in the training examples of the deep learning model. Although these frameworks successfully improved the accuracy of predictions, despite their computationally intensive training procedure, they fall short when it comes to generalizing to new environments. For example, as attested in \cite{BLSTM}, the accuracy of the model reduces to 32\% when tested in an untrained environment. 
The authors in \cite{generalizeAppDependent} suggested extracting the power distribution of different gestures' velocities from the Doppler spectrum. Then, a temporal learning model is applied to learn the extracted features and perform domain-independent gesture recognition.
However, these approaches require extensive data pre-processing, careful feature extraction, and an abundance of training examples. For example, experimental results in \cite{generalizeAppDependent} show that the accuracy of the algorithm drops significantly when the number of human subjects and training samples decreases. 

Conversely, \gls{fsl} is a novel \gls{ml} technique that has shown remarkable results in generalizing from a few examples and classifying examples from previously unseen classes given only a handful of training data. Thanks to this unique feature, \gls{fsl} can be an exceptional candidate for \gls{csi} learning problems. A recent work \cite{harOneshot} addresses human activity recognition through \gls{fsl}. A \gls{csi} feature extraction method along with a \gls{matnet} \cite{matchingNet} is used to remove the environment-dependent data and to make accurate predictions in new environments. However, \glspl{matnet} require designing application-dependent embedding functions, through \gls{lstm} and attention-based \gls{lstm} networks. This adds to the computational complexity of the algorithm and limits its generalizability to only certain applications. 
\textit{To address these challenges, we use state-of-the-art \gls{prnet} \cite{snell2017prototypical}, which (i) achieves the same level of performance with a simpler learning structure, and (ii) can be generalized to a variety of \gls{csi} sensing applications.} \vspace{-0.1cm}

\subsection{CSI Collection and Datasets Availability}\label{sec:csi_collection}

Reliable and high-resolution \gls{csi} datasets are key enablers of any learning-based \gls{csi} sensing applications. In Wi-Fi, \gls{csi} data can be estimated at the receiver over all subcarriers through pilot symbols contained in the \gls{phy} preamble. Being computed at the \gls{phy}, \gls{csi} is not accessible by the end-user through normal \glspl{nic}, which makes \gls{csi} acquisition a challenging task. Some \gls{sdr} Wi-Fi implementations do exist \cite{bloessl2013ieee}, but to the best of our knowledge, they are limited to 20 MHz bandwidth only (IEEE 802.11 a/g/p).  For this reason, the majority of research works have used 802.11g and 802.11n standards for data collection, which limits the total bandwidth to 40 MHz \cite{Intelchip}. To address this shortcoming, we leverage the recently released Nexmon CSI-extractor tool \cite{nexmonCSI} to obtain CSI from IEEE 802.11ac transmissions at 80 MHz of bandwidth. By leveraging \gls{mimo}, we collected CSI with four receiver antennas. In Section \ref{sec:perf_eva}, we show that increasing the number of antennas, receivers, and subcarriers over which the \gls{csi} data is measured increases the \gls{csi} sensing performance significantly.

To the best of our knowledge, only a few \gls{csi} sensing datasets are publicly available. FalldeFi \cite{falldefi} is a fall detection dataset consisting of information collected in 6 rooms. \gls{har} dataset is collected in \cite{yousefiData} and \cite{guo2019wiar} where measurements are collected inside 1 and 3 rooms, respectively. In all cases, 1 transmitter and 1 receiver are utilized. The size of each instance in all datasets is (2000, 3, 30), representing 2000 \gls{csi} matrices with three receive antennas and 30 subcarriers measured in two seconds. In this paper, for the first time, \textit{we collect  a large-scale dataset containing multi-antenna (up to 4) multi-receiver (up to 3) fine-grained (up to 242 subcarriers) \gls{csi} readings from multiple environments and activities, which we will release to the community.}
 
\section{The \FW Framework}\label{sec:ProposedScheme}

This section describes the proposed \FW framework for robust \gls{csi} sensing. Our framework's core focus is on (i) reducing the cost of data acquisition and labeling by adopting customized \gls{prnet} to learn and generalize with limited data; (ii) reducing the complexity of the learning algorithm and data pre-processing while maintaining the accuracy by collecting data over multiple antennas, multiple receivers at different locations and high-resolution \gls{csi} data over large channel bandwidth. We first explain the core design principles in Section \ref{sec:rob_princ}. Then in Section \ref{sec:csi_fusion}, we describe  the \FW \gls{csi} processing procedure.  Finally,  we present our \gls{fsl}-based technique for robust CSI learning in Section \ref{sec:csi_learning}. \vspace{-0.1cm}

\subsection{\FW Robust Design Principles}\label{sec:rob_princ}

The \FW is a passive sensing system leveraging \gls{csi} data computed through listening to ongoing traffic exchange between  Wi-Fi devices. The key motivation behind the \FW is that the accuracy of the \gls{csi} computation depends on the coherence time of the channel, the received power of the transmitted signal, interference from other transmissions and background noise, among other factors. For this reason, the \FW leverages multiple receivers and multiple antennas to collect robust \gls{csi} data and thus increases the accuracy. Below, we detail the impact of each factor in increasing the accuracy of \gls{csi} sensing. \smallskip

\textbf{(1) Spatial diversity.}~Due to the presence of multiple reflectors and scatterers in indoor environments, there is a significant probability that the communication channel is in a deep fade, and thus, that the CSI measurement may be erroneous. This motivated us to increase the reliability of the \gls{csi} sensing framework through adding spatial diversity. Since we do not have any control over the final configuration of \glspl{ap} in the target environment, we must ensure that the \gls{ofdm} symbols for \gls{csi} estimation pass through multiple signal paths, each of which fades independently, guaranteeing that reliable \gls{csi} measurement is possible even if some paths are in a deep fade. Therefore, in \FW we incorporated two types of diversity, namely, \textit{macro-diversity} and \textit{micro-diversity}.

\textbf{(a)} \emph{Micro-diversity.}~\FW improves performance by collecting CSI data with multiple receive antennas. As long as the antennas are placed sufficiently far apart, this would create independent propagation paths between different antenna pairs, thus reducing the chance of deep fade significantly. In indoor environments, the channel decorrelates over shorter spatial distances \cite{tse2005fundamentals}, and the typical Wi-Fi router's antenna separation of the half to one carrier wavelength is sufficient (about 3-6 cm at 5GHz band). To make \FW compatible with legacy Wi-Fi systems, we chose to leverage one spatial stream and use the full diversity on the receiving side by activating four receive antennas. \textbf{Our experimental results in Table \ref{tab:results} show that micro-diversity helps increase classification accuracy by up to 16\% with respect to a single-antenna system}, as micro-diversity improves robustness to multipath fading and interference. 

\textbf{(b)} \emph{Macro-diversity.}~\FW leverages the usage of multiple receivers to further improve diversity. Indeed, modern indoor Wi-Fi networks commonly leverage multiple \glspl{ap} to increase the wireless range, boost coverage, ensure reliability and support a large number of wireless applications/devices. Thus, \FW collects \gls{csi} simultaneously from multiple receivers. \textbf{The experimental results in Table \ref{tab:results} show that macro-diversity improves the prediction accuracy by 38\% with respect to a single-receiver system}, as the receivers are located several wavelengths apart from each other. On the other hand, macro-diversity implies that the data from multiple receivers must be aligned both in time and frequency domains.

\textbf{(2) Time diversity.}~The Wi-Fi propagation environment is subject to almost continuous change, mainly owing to the movement of obstacles between the transmitter and the receiver, as well as the presence of noise and interference from overlapping channels. For this reason, to improve robustness, \FW leverages the usage of multiple, subsequent CSI readings to boost the classification accuracy. To compensate for the increase in complexity, we use a novel and custom-tailored technique based on singular value decomposition (SVD), which is explained in Section \ref{sec:csi_fusion} (Step 4: Dimension reduction). \textbf{We show that our technique helps reduce the input size by about 80\%.}  Moreover, by trading off delay for accuracy (the more CSI readings, the more delay), \textbf{we show in Figure \ref{fig:csi_fusion} that time diversity increases accuracy by up to 35\%}, as the number of CSI readings fed to the learning model help counteract the adverse channel conditions. 

\textbf{(3) Subcarrier resolution.}~\gls{csi} measurement at low channel bandwidth cannot resolve multipath propagations and thus limiting many \gls{csi} sensing applications that require higher precision \cite{wifiRespiratoryWide,tonetrack}.  Inevitably, either feature extraction algorithms have been utilized or sampling frequency has been increased, to compensate for the low resolution in the frequency domain. In order to avoid time- and computational- intensive data pre-processing for complex feature extraction and efficient sampling, we proposed using the Nexmon tool \cite{nexmonCSI} to extract high-resolution 802.11ac \gls{csi} data over the 80 MHz wide channel with 242 subcarriers. \textbf{The results in Figures \ref{fig:resolution_1rx} and \ref{fig:resolution_3rx} show that using higher subcarrier resolution improves prediction accuracy by up to 19\%}, as finer-grained CSI data is fed to the learning model.\vspace{-0.1cm}

\subsection{\FW CSI Processing} \label{sec:csi_fusion}

\begin{figure}[h!]
    \centering
    \includegraphics[width=1\columnwidth]{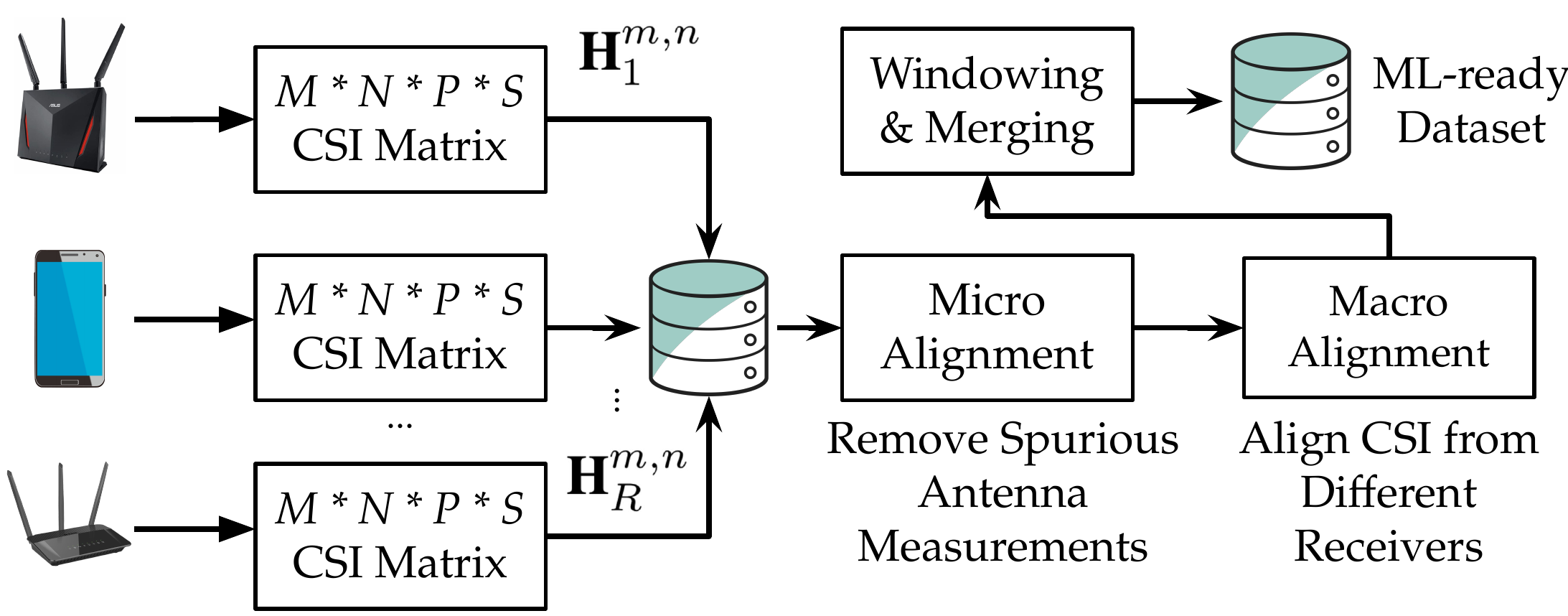}
    \caption{\FW CSI dataset preparation and dataset construction.\vspace{-0.3cm}}
    \label{fig:csi_fusion}
\end{figure}

We now describe the procedure \FW converts the unprocessed \gls{csi} measurements into a dataset used as the input of the \gls{ml} algorithm. We term this
processing \textit{\gls{csi} data preparation}, and is depicted in Figure \ref{fig:csi_fusion}.

As explained earlier, \FW leverages multiple receivers with multiple antennas, as well as fine-grained \gls{csi} estimated by the Wi-Fi \gls{ofdm} receiver. In an \gls{ofdm} system, the digital data stream is modulated over multiple overlapping, closely spaced orthogonal subcarriers to transmit data in parallel. Let us assume $P$ packets are captured during the data collection campaign. In a $M \times N$ \gls{mimo} \gls{ofdm} system with $M$ transmit antennas, $N$ receive antennas and $S$ subcarriers, the extracted \gls{csi} matrix between the transmit antenna $m$ and the receive antenna $n$ located on receiver $r$ can be written as \[
\mathbf{H}_r^{m,n} = \begin{bmatrix} 
    h_{1,1}^{m,n} & \dots & h_{1,s}^{m,n}& \dots& h_{1,S}^{m,n} \\
    \vdots & & \vdots& &\vdots\\
    h_{p,1}^{m,n} & \dots & h_{p,s}^{m,n}& \dots &h_{p,S}^{m,n} \\
    \vdots & \ddots & \vdots&\ddots &\vdots\\
    h_{P,1}^{m,n} & \dots & h_{P,s}^{m,n}& \dots &h_{P,S}^{m,n} 
    \end{bmatrix}, \begin{tabular}{c} $1 \le n \le N$ \\ $1 \le m \le     M$,\end{tabular}
\]
where the element $h_{p,s}^{m,n}$ denotes the amplitude and phase information of the \gls{csi} obtained from $p$-th packet, the $s$-th \gls{ofdm} subcarrier over the channel from the transmitter $m$ to the receiver $n$. For example, when $P=100$ and 80 MHz channels, the matrices $\mathbf{H}_r^{m,n}$ have $P \times S = 100 \times 242$ elements. 

\textbf{Step 1. Micro alignment.}~During our experiments, we noticed that in some instances, \gls{csi} packets are not captured by all antennas. In that case, those \gls{csi} measurements are identified and removed. 

\textbf{Step 2. Macro alignment.}~In addition to antenna-level alignment, the data collected from different receivers are aligned to ensure that each \gls{csi} element collected over different receivers represents the same time and frequency domain channel measurements. To this end, the Wi-Fi sequence number is used to match data from different receivers. 

\textbf{Step 3. Segmentation and integration.} At this step, first, to remove noise and unwanted amplification, the \gls{csi} elements are normalized by the mean amplitude over all subcarriers. 
Next, the \gls{csi} measurements from each antenna, $\mathbf{H}_r^{m,n}$ is divided into fixed-size \textit{data-segments} by sliding a non-overlapping window through the time-domain measurements.  The data-segments, denoted by $\mathbf{\hat{H}}_{r}^{m,n}$, are matrices of dimension $W \times S$, where $W$ is the number of \gls{ofdm} packets in a window. Further, measurements from all antennas on the same receiver are stacked into one matrix to form a \textit{data-frame} as
\begin{equation}
  \mathbf{H}_{r} = 
  \left[
        \hat{\mathbf{H}}_r^{m,1},  
        \cdots,
        \hat{\mathbf{H}}_r^{m,N}
    \right]^T\label{eq:data_frame}
\end{equation}
where data-frame $\mathbf{H}_{r} = \mathbf{H}_{r}^A e^{j \mathbf{H}_{r}^{\phi}}$ is a complex matrix. Note that the transmit antenna index $m$ is omitted in the interest of readability. Moreover, both amplitude matrix $\mathbf{H}_{r}^A = \|\mathbf{H}_{r} \|$ and phase matrix $\mathbf{H}_{r}^{\phi} = \measuredangle \mathbf{H}_{r}$ can be utilized for \gls{csi} sensing individually. For notation brevity, from here on, $\mathbf{H}_{r}$ is a real matrix representing both $\mathbf{H}_{r}^A$ and $\mathbf{H}_{r}^{\phi}$ matrices.\smallskip

\textbf{Step 4. Dimension reduction.}~Data-frames with size $N\times W \times S$ may be too large to be fed into the  learning module. To improve the performance and processing time of the learning algorithm, data-frames are minimally pre-processed with \gls{svd}, which is a powerful tool to eliminate the less important variables of large-size data matrices and produce an approximation with lower dimensions. To preserve the subcarrier resolution, we only reduce the number of features in the time domain, i.e., \gls{csi} packets. By using \gls{svd}, the data-frame matrix in \eqref{eq:data_frame} is factored into the product of three matrices as $\mathbf{H}^T_{r} = U\Sigma V^T$, where diagonal values of $\Sigma$ contains the singular values of the data-frame and $U$ and $V$ are known as the left- and right-singular vectors. 

The key to understanding \gls{svd} functionality is that by multiplying $\mathbf{H}_r^T$ with the left-singular vector $V$, time-domain measurements over each subcarrier is mapped to the \textit{subcarrier space}, to preserve only the useful packets over each subcarrier. Using eigenvalue analysis, we noticed that singular value contributions of the data-frames depend on the type of activity and the environment. Hence, we utilize all singular values to maintain the \FW generalizability. The compact data-frame, $\mathbf{H}'$ with dimension $S \times S$ is calculated as 
\begin{equation}
    \mathbf{H}'_r = \mathbf{H}^T_{r} \times V.
\end{equation}
Further, to extract more features from the compact data-frame, 
the linear correlation among subcarriers is extracted simply by calculating the Pearson Correlation Coefficient (PCC) of the compact data-frame \cite{han2011data}. These correlation matrices, denoted as $\rho_{\mathbf{H}}$ with dimension $S \times S$ are the input of the \FW learning module. Therefore, we are able to reduce the dimensionality from $N\times W \times S$ to $S \times S$, \textit{which makes the input constant with the window size and the number of receive antennas}. For example, since in our experiments we used 4 antennas and a window size of up to 300, we are able to reduce the input size \textit{by about 80\%}.

\subsection{\FW Learning Models}\label{sec:learning}

We first discuss preliminaries about the \gls{fsl} and related models. Then, we present the \FW CSI learning procedure.

\textbf{\gls{fsl} and \gls{prnet}.}~Traditionally, supervised learning approaches require a large labeled dataset for training. In applications such as Wi-Fi sensing, collecting and labeling large datasets may be challenging, thus the labeled data is severely limited \cite{ganHAR}. For this reason, \FW uses few-shot learning (FSL), where the objective is to quickly adapt to new/unseen data given a limited number of samples. The purpose of \gls{fsl} is to train a model with high accuracy when the data of the target task is small by using some kind of prior knowledge. In practice, \gls{fsl} is useful when training examples are hard to find, or the cost of labeling is high \cite{matchingNet}. 

\begin{figure}[h]
    \begin{subfigure}[t]{0.32\columnwidth}
    \centering
    \includegraphics[width=1\columnwidth]{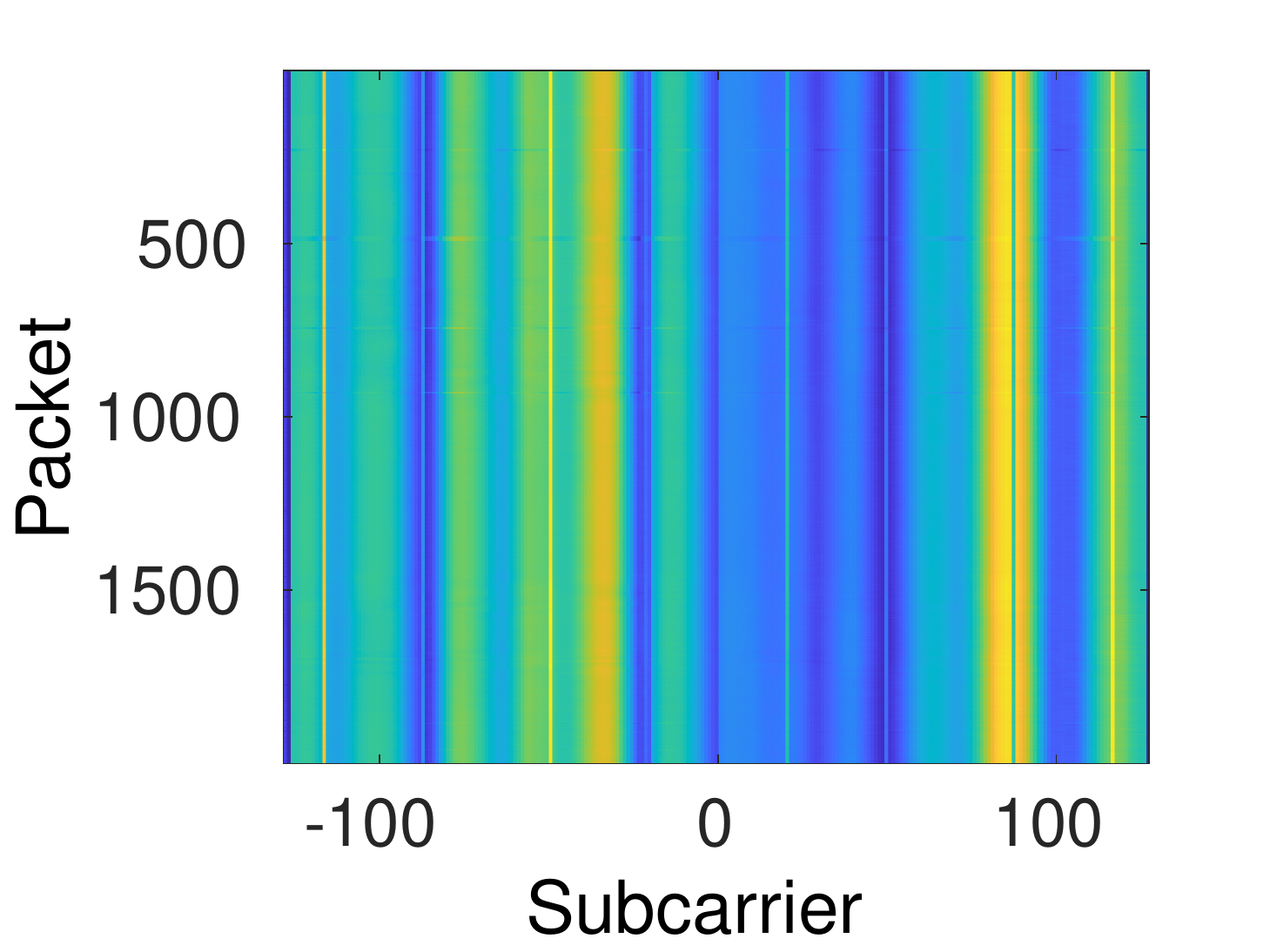}
    \caption{$E_1$}
    \end{subfigure}
    \hfill
    \begin{subfigure}[t]{0.32\columnwidth}
    \centering
    \includegraphics[width=1\columnwidth]{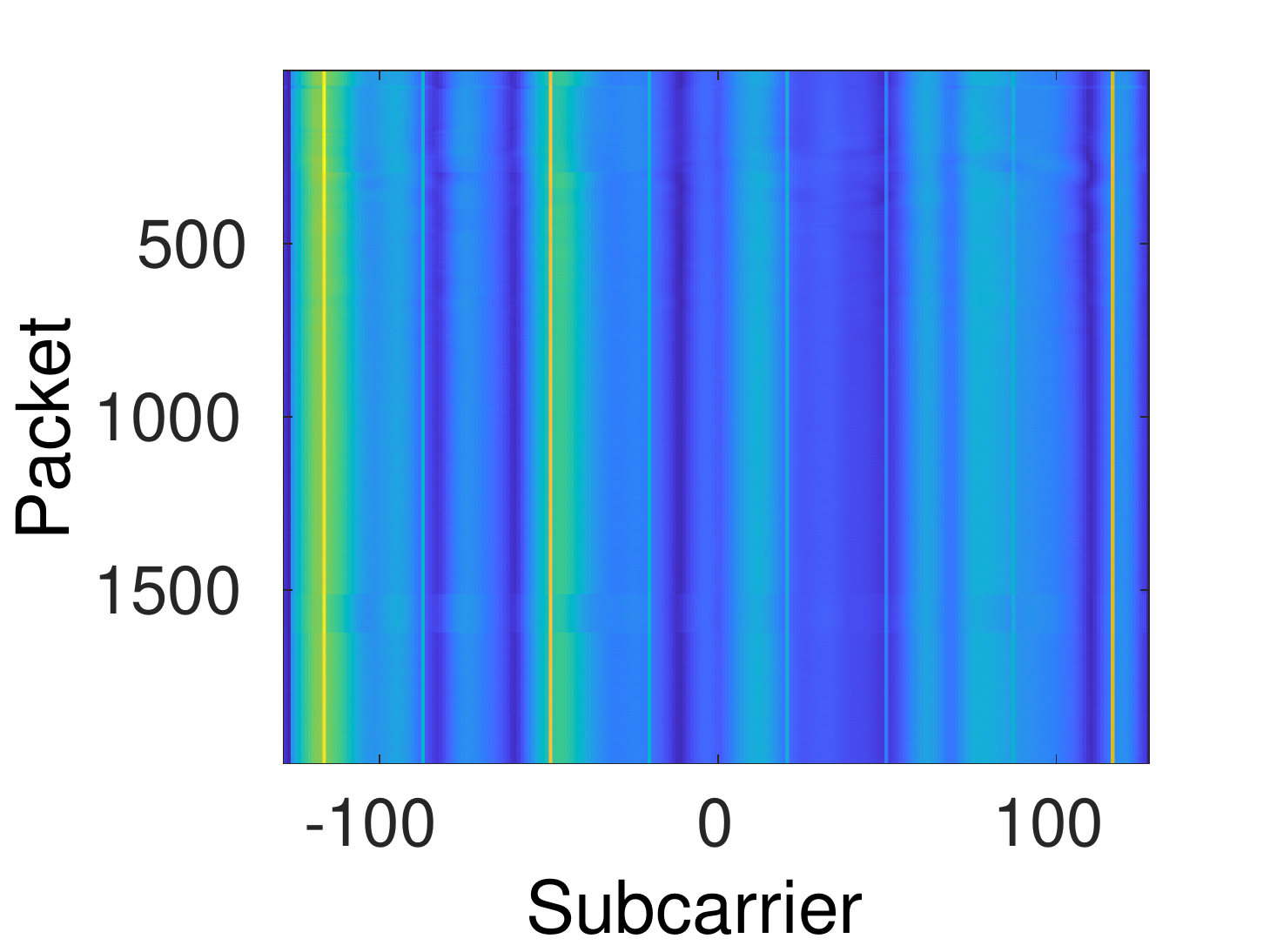}
    \caption{$E_2$}
    \end{subfigure}
    \hfill
    \begin{subfigure}[t]{0.32\columnwidth}
    \centering
    \includegraphics[width=1\columnwidth]{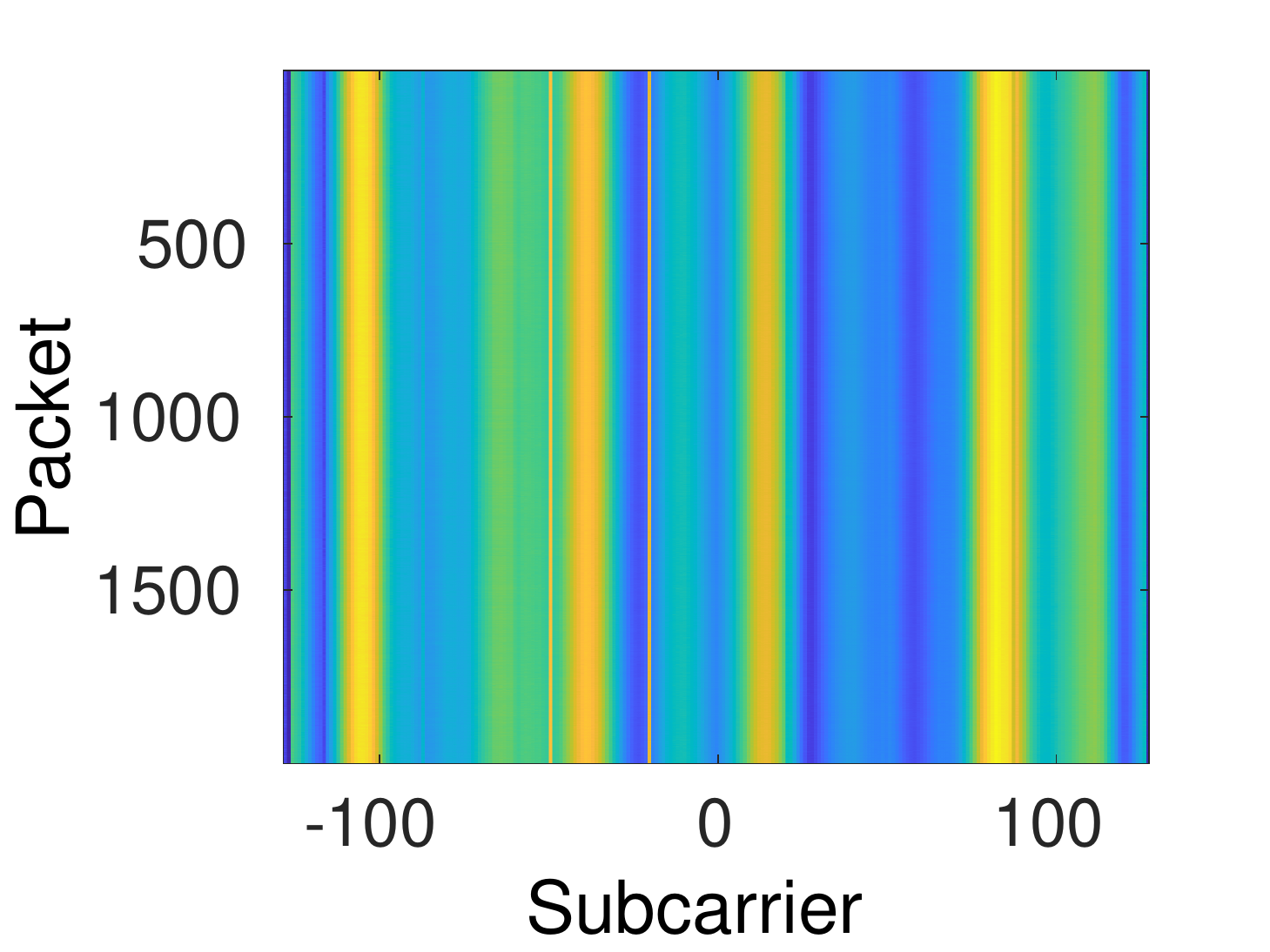}
    \caption{$E_3$}
    \end{subfigure}
    \caption{\gls{csi} measurements of an empty room in the data collection environments described in Figure \ref{fig:environments}.\vspace{-0.25cm}}
        \label{fig:empty_comparison}
\end{figure}

Specifically, in $K$-way-$N$-shot learning, the model is trained through a set of sampled mini-batches of classes and data points, called \textit{training tasks}. Each task is divided into a query set and limited support set with $K$ different classes and $N$ labeled examples (shots) of each class (typically 1-5). The small training set requires the fast adaptability of the model. The query set is classified using the knowledge from the support set. The performance of the model in generalizing to new classes is evaluated by the average test accuracy across many $K$-way-$N$-shot \textit{test tasks} containing unseen/new classes.
In a nutshell, by learning a generalizable metric space, \gls{fsl} learns \textit{how} to classify given a set of training tasks, and exploit this knowledge to classify new classes. It is worth noting that in \gls{csi} sensing applications a new environment can be interpreted as a new class. To illustrate this point, Figure \ref{fig:empty_comparison} shows \gls{csi} measurements of empty rooms in 3 different environments. \textit{Although all three are labeled as \textit{empty room}, it can be seen that they show distinct patterns due to the unique propagation profile of each environment}.

Different \gls{fsl} approaches, such as \gls{matnet} \cite{matchingNet}, meta-learning \cite{metaLearning} and \gls{prnet} \cite{snell2017prototypical}, propose different metric spaces in which classification can be performed. Among these methods, \gls{prnet} applies a simple yet effective inductive bias in the form of class \textit{prototypes} that leads to achieving impressive few-shot performance and reducing the network complexity. Next, we briefly explain the fundamentals of the \glspl{prnet}.\smallskip 

\textbf{\gls{prnet} model.}~
During $e$-th training \textit{episode}, a \textit{training task} comprising a mini-batch of classes and data points are sampled as 
$\mathcal{T}_e = \{ \mathcal{D}_1, \ldots, \mathcal{D}_K \}$. $\mathcal{D}_k = \{(\mathbf{x}_i, y_i) | y_i = k\}$ denotes the set of data points labeled with class $k$, where each $\mathbf{x}_i$ is a data point and $y_i \in \{1, \ldots, K\}$ is the corresponding label. Next, $\mathcal{T}_e$ is further divided into \textit{support} and \textit{query} sets, i.e., $\mathcal{T}_e = \{\mathcal{S}_e,\mathcal{Q}_e\}$.
A subset of each class set, $\mathcal{D}_k$ with $N$ instances is selected as the support set: $\mathcal{S}_e = \{ (\mathbf{s}_1, y_1), \ldots, (\mathbf{s}_N, y_1), \ldots,(\mathbf{s}_N, y_K) \}$ and the rest of the examples are used as the query set, $\mathcal{Q}_e= \{ (\mathbf{q}_1, y_1), \ldots, (\mathbf{q}_L, y_1), \ldots,(\mathbf{q}_L, y_K) \}$ where $L$ is the number of samples in the query set. 

\begin{figure}[h!]
    \centering
    \includegraphics[width=.9\columnwidth]{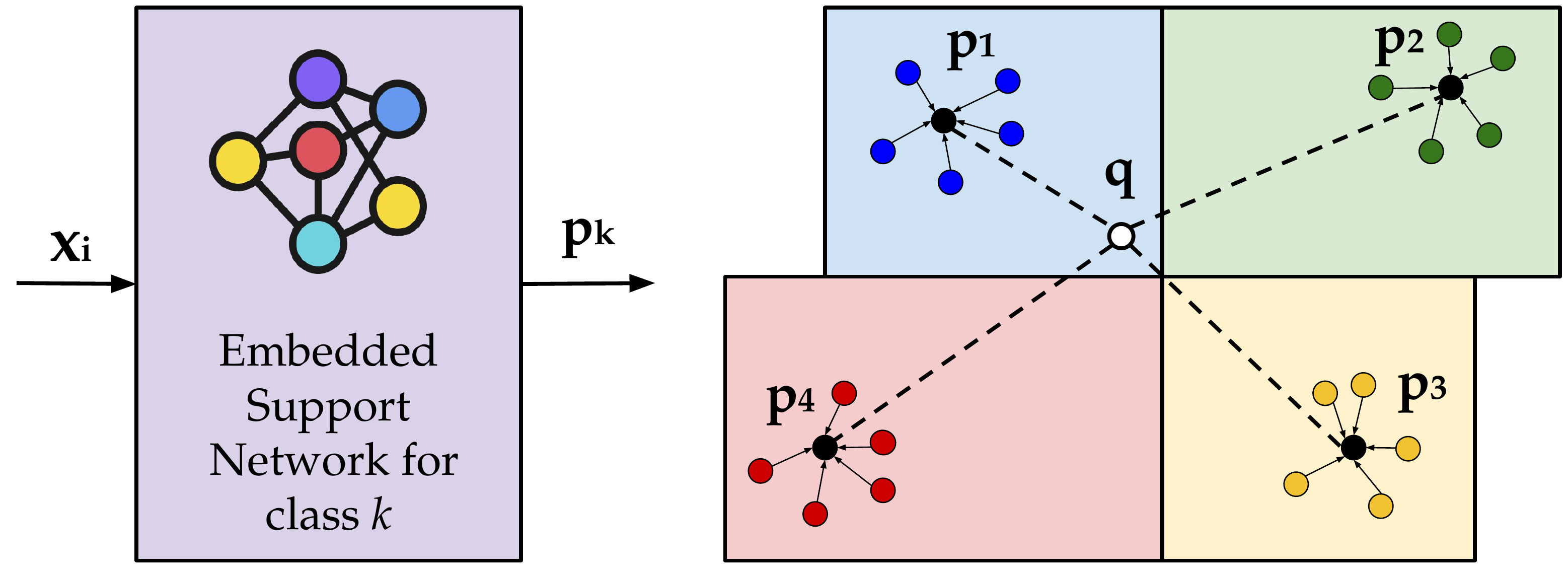}
    \caption{Embedded Prototype Network.\vspace{-0.25cm}}
    \label{fig:protonet}
\end{figure}

To avoid suffering from high variance caused by the high dimensionality of the $\mathbf{x}_i$, the support and query data are mapped into a feature space by embedding function $f_{\theta}(\mathbf{x}_i)$ with cleanable parameters $\theta$, as shown in Figure \ref{fig:protonet}. Learning a proper embedding model is a paramount task \cite{rethinkingEmbedding} that will be discussed later in-depth in this section.

The principle idea behind \gls{prnet} is that data points cluster around a single prototype representation for each class that is simply the mean of the embedded support samples of each class \cite{snell2017prototypical}. Therefore,  the \textit{prototype} of each class, $\mathbf{p}_k$, is computed as
\begin{equation}
    \mathbf{p}_k = \frac{1}{|\mathcal{D}_k|} \cdot \sum_{(\mathbf{s}_i, y_k) \in D_k} f_{\bm{\phi}}(\mathbf{s}_i)
   \label{eq:prototype}
\end{equation}
These prototypes can be used to classify the query samples, as illustrated in Figure \ref{fig:protonet}. For a query point $\mathbf{q}$, the ProtoNet produces a distribution over classes using a softmax over distances to the prototypes in the embedding space as
%
\begin{equation}
\begin{split}
    \mathcal{L}(\mathcal{Q}_e) =  \frac{-1}{|\mathcal{Q}_e|}\sum_{(\mathbf{q}_i,y_i)\in \mathcal{Q}_e}\log \left(\frac{\exp(-\|f_{\bm \theta}(\mathbf{q}_i)- \mathbf{p}_k)\|^2)}{\sum_{k'} \exp(-\|f_{\bm \theta}(\mathbf{q}_i)- \mathbf{p}_{k'}\|^2)}\right). \label{eq:loss_proto}
\end{split}
\end{equation}
%
The \gls{prnet} is trained through minimizing the loss function via \gls{sgd} over training episodes. At test time, sample $\textbf{x}_i$ is classified using the nearest-neighbouring prototype computed from the support set of the test episode as  $y(\mathbf{x}_i) = \arg\min_{j \in \{1,...,K\}}\|\mathbf{x}_i-\mathbf{p}_j\|^2$.
It is worth noting that \FW adopts Euclidean distance rather than cosine similarity used in \cite{harOneshot} as it is shown through extensive experiments that it outperforms other distance metrics \cite{snell2017prototypical}.\smallskip

\textbf{Learning embedding model $f_{\theta}$.} The \gls{prnet}'s goal is to learn a transferable embedding that generalizes to new tasks. Previous work \cite{harOneshot} has adopted a \gls{matnet} and employed two distinct \glspl{dnn} for query samples and support samples. In addition, attention-based \gls{lstm} is used to encode a full context embedding of support samples. \textit{Unlike \cite{harOneshot}, we use only one embedding function for both support and query sets} rather than multiple embedding functions. \textit{This reduces the number of hyper-parameters, simplifies the learning process of \gls{csi} sensing and reduces the time- and computational- complexity.} In addition, learning the embeddings according to the dataset \textit{makes \FW application-independent}. 
Inspired by \cite{rethinkingEmbedding}, \FW learns the embedding function $f_{\theta}$ by training a neural network on the entire training set, i.e., we merge all the support mini-batches into a single set as $\mathcal{S} = \cup \{\mathcal{S}_1, \ldots, \mathcal{S}_e, \ldots, \mathcal{S}_E\}$, where $E$ is the total number of training tasks. The embedding model parameters is then achieved by 
\begin{equation}
        \theta = \arg\min_\theta \mathcal{L}^{ce} (\mathcal{S}; \theta),
\end{equation}
where $\mathcal{L}^{ce}$ denotes the cross-entropy loss between predictions and ground-truth labels.

\subsection{\FW Training and Inference}\label{sec:inference}

\FW learns from the training dataset collected in the \textit{source} environment to perform the related \gls{csi} sensing application in the \textit{target} environment. 
The dataset of each receiver $r$ consists of \gls{csi} data-frames of size $S \times S$ and corresponding labels, as explained in Section \ref{sec:csi_fusion}. Using these datasets, a learning model is trained for each receiver. 
The training dataset is sampled into mini-batches of training and testing tasks. Each task includes a support and a query set of $K$ classes and $N$ examples (shots) as explained in Section \ref{sec:learning}.\smallskip

\textbf{Training.}~First, the support mini-batches are fed into an embedding learning module to train a proper embedding function. The embedding is trained using a \gls{cnn} with four convolutional blocks. Each block comprises a 64-filter $3 \times 3$ convolution, batch normalization layer \cite{batchNormalization}, a ReLU nonlinearity, and a $2 \times 2$  max-pooling layer is applied after each of the blocks. Finally and a global average-pooling layer is on top of the fourth block to generate the feature embedding. All of our models were trained via \gls{sgd} optimizer with Adam \cite{Adam}. The learning rate is initialized as $10^{-3}$ and is cut in half every 2000 episodes.  Following learning the proper embedding model, the learning proceeds by minimizing the loss function in \eqref{eq:loss_proto}. \smallskip

\textbf{Testing.}~During testing time, the algorithm only requires $N$ samples (also called \textit{shots}) of the desired activities to function in the target environment. In a real-world scenario, this can be realized through a mobile application installed on the end-users smartphone.  As a part of the initialization process, the user is asked to perform some activities for a short amount of time (less than a minute) and label the activities. The learning algorithm utilizes the provided labeled data by the user as the support set of the testing task and adapts to the target environment. Note that initialization and collecting data in the target environment is a common practice of commercial activity recognition tools, such as Qualcomm's positioning units \cite{karmanov2021wicluster}, and does not impact the practicality of this approach. As explained earlier, there exists a trained model per receiver. At testing time, each model returns a probability distribution vector over all classes as $P_r = [p_1,\ldots, p_k]$. By defining $R$ as the number of receivers, the final decision on the label of a data point $y$ is made through the superposition of all probability vectors and finding the maxima 
$y = \arg\max_k \sum_{r =1 }^{R} P_r$.

\section{Prototype and Experimental Data Collection}\label{sec:exp_data_coll}

To demonstrate the robustness of \FW in generalizing to new environments, we designed a testbed with commercially-available off-the-shelf Wi-Fi devices. We evaluate the performance of \FW by building a prototype and analyzing a use-case application of \gls{csi} sensing, i.e., activity recognition. First, we present our evaluation methodology including environment setup, measurement tools, data collection campaigns, and training/testing procedure. Then, we show our results which investigate the following crucial aspects:
\begin{enumerate}
    \item The role of \gls{prnet} in enabling \FW generalizing to unseen environments.
    \item The impact of macro-, micro- diversity, and subcarrier resolution on the accuracy of \FW predictions.
\end{enumerate}
Note that our main goal is to demonstrate the ability of \FW in generalizing to new environments, not recognizing human activities. Thus, we suffice our experiment to only four activities, namely \textit{empty room}, \textit{walking}, \textit{jumping}, and \textit{standing}, each of which has its own specific challenges. Notice that even identifying an empty room is not a trivial task, as changes in environment/time of experiment lead to significant alterations in the \gls{csi} measurements as shown in Figure \ref{fig:empty_comparison}. Considering various activities with more complexities will be left to our future work.
The implementation of the \FW prototype requires a pair of Wi-Fi transmitter-receiver to establish a traffic link, in addition to a set of Wi-Fi routers equipped with an extraction tool to process the \gls{csi} data. In the following, we detail the \FW components as well as the experiment setup.\vspace{-0.2cm}

\subsection{CSI Extraction, Hardware and Testbed Setup}

We have used Nexmon \gls{csi} \cite{nexmonCSI}, the state-of-the-art \gls{csi} extraction tool to collect \gls{csi} measurements using Asus RT-AC86U Wi-Fi routers.  In a real-world scenario, \FW relies on already existing Wi-Fi transmissions. In our controlled experiment, we dedicated a pair of transmitter-receiver to emulate  the traffic generation. To establish the Wi-Fi link, a Netgear R7800 Wi-Fi router with a Qualcomm Atheros chipset is used in \gls{ap} mode. An off-the-shelf laptop acts as the client. The Hostapd tool \cite{rosen2009linux} is utilized on the \gls{ap} to ensure that the traffic is generated using 802.11ac at the desired bandwidth. UDP packets between the \gls{ap} and the client are generated with a rate of 1 Mbit/s via iperf3 tool \cite{mortimer2018iperf3} and transmitted through a single spatial stream. 

To implement our prototype \FW, we have used 3 Asus RT-AC86U WiFi routers, each equipped with $N=4$ antennas.  The Asus routers extract the \gls{csi} packets using the Nexmon firmware, by computing the \gls{csi} on the UDP frames transmitted from the \gls{ap} to the client. \gls{csi} is computed at a rate of 100 Hz. The Nexmon tool enabled us to collect 802.11ac channel measurements at 5GHz with 20 and 80 MHz bandwidth over $S = 52$ and $S = 242$ subcarriers. The measurements at 20 MHz are used to evaluate the effectiveness of higher subcarrier resolutions as compared with legacy \gls{csi} measurements. Note that originally 802.11ac 20 MHz and 80 MHz channels consist of 64 and 256 subcarriers. However, in our measurements, we discarded the \gls{csi} from the guard and null subcarriers as they contain arbitrary values \cite{nexmonCSI}.

To evaluate the capability of \FW to generalize to different environments, we performed \gls{csi} measurements in three different environments. Specifically, the measurements are carried in an office area $E_1$, a meeting room $E_2$, and a classroom $E_3$ on different days and times. We carefully picked the environment with exclusive furniture arrangements, size, and construction material to ensure that the target environments are mutually exclusive from the source environment in terms of propagation characteristics. However, visually, $E_1$ is more similar to $E_3$ than $E_2$. Figure \ref{fig:environments} shows the environment layout, as well as the position of \gls{ap}, client and \gls{csi} sniffers. The position of \FW components may influence the test outcome. Therefore, we attempted to loosely change the configuration in different environments and design the testbed to align with realistic home/office \gls{har} scenarios and be close to those used in other CSI activity recognition studies \cite{ganHAR, wifiHAR}. 
\smallskip
\subsection{Data Collection and Dataset Preparation}
\begin{figure}[!ht]
    \centering
    \includegraphics[width=.85\columnwidth]{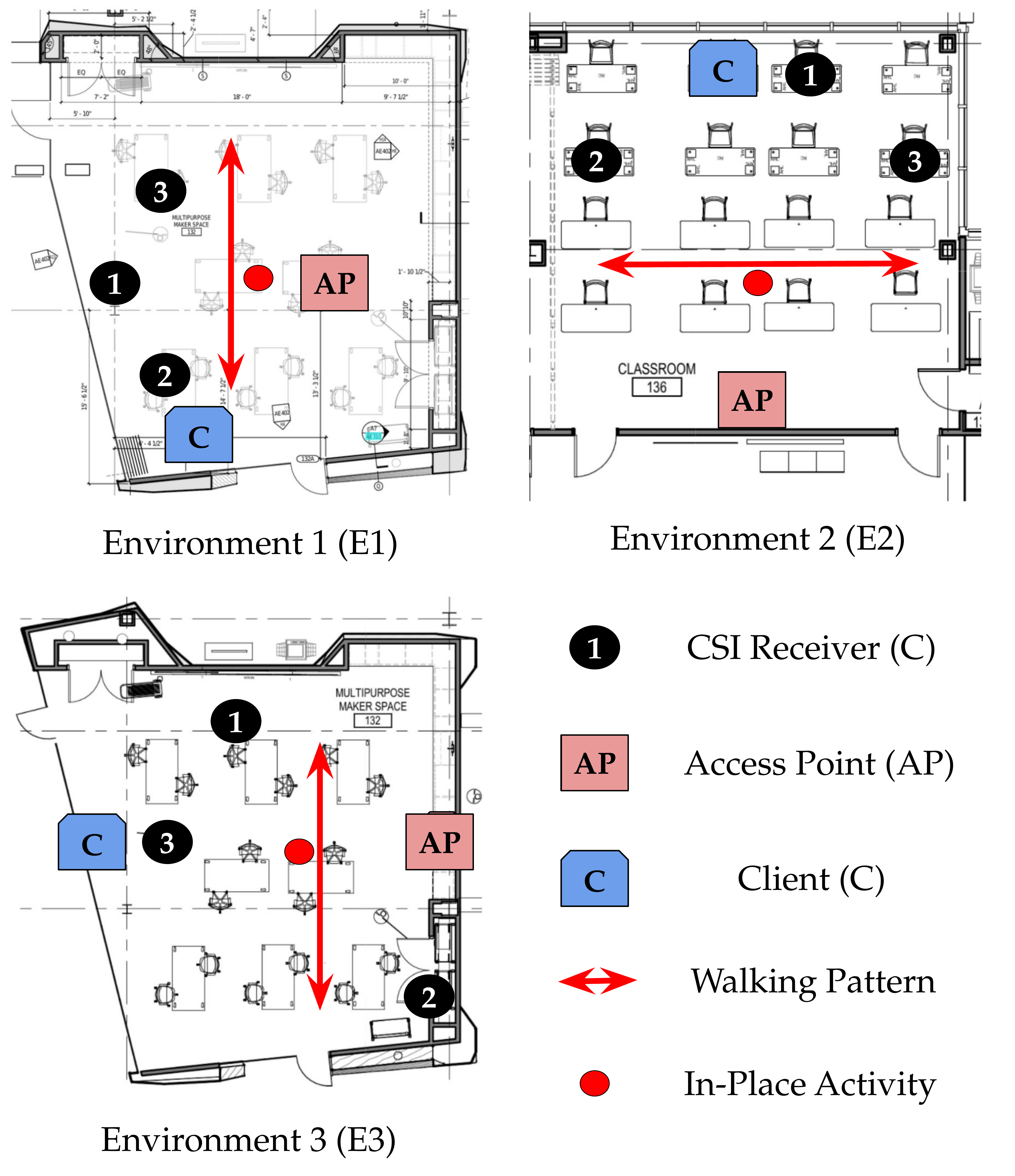}
    \caption{Data collection and testing environments used in our experiments. We also report the legend of symbols used in the figure.\vspace{-0.4cm}}
    \label{fig:environments}
\end{figure}

The two subjects involved in the experiments were instructed about the type, duration, and location of the activities including jumping, walking, and standing. Each measurement campaign involves 180 seconds of data collection for each activity performed by two people (IRB approval available upon request). The measurements are repeated 10 times with a time interval of at least 2 hours in between measurements. The collected raw data is processed by applying the \gls{csi} data preparation algorithm presented in Section \ref{sec:csi_fusion}. Upon aligning the data collected from 4 antennas of each Asus router, a window size of $W = 300$ samples (tantamount to 3s) is used to segment the raw data into data-segments. Further, the data-segments are integrated using \eqref{eq:data_frame} to form the \gls{csi} data-frame $\mathbf{H}_{r}$. As a result, through the explained measurement campaign with $M = 1$  spatial stream, $N = 4$ receive antennas, the \gls{csi} data-frame $\mathbf{H}_{r}$ at 80 MHz is a matrix of size $1200 \times 242$. Thanks to the dimension reduction algorithm proposed in Section \ref{sec:csi_fusion}, the size of the data-frame is reduced to $242 \times 242$, which is about 80\%. We collected data in three different propagation environments, which are shown in Figure \ref{fig:environments}.

Environment $E_1$ is selected as the source environment, while $E_2$ and $E_3$ are considered as the target environments, and their dataset is entirely used for testing purposes. The \FW learning algorithm presented in Section \ref{sec:learning} has been trained using the prepared dataset, where 70\% of the dataset is used for training while the remainder is used for testing and evaluation. Unless otherwise mentioned, the model is trained using 4-way-5-shot training tasks, i.e., 4 classes and 5 examples in both support and query tasks. The accuracy of the algorithm is tested through sampling 1000 randomly generated testing tasks from the test sets.
We have used two measures to report the results, namely accuracy and F1 score. The accuracy refers to the correct predictions divided by total predictions, while the F1 score is defined as 
\begin{equation}
    \text{F1 score} = 2\times \frac{\text{Precision}\cdot\text{Recall}}{\text{Precision}+\text{Recall}}\nonumber
\end{equation}
where $\text{Recall}\coloneqq \text{TP}/(\text{TP}+\text{TN})$ and $\text{Precision}\coloneqq \text{TP}/(\text{TP}+\text{FP}$). $\text{TP}$, $\text{TN}$ and $\text{FP}$ stand for true positive, true negative and false positive, respectively.

\begin{table*}
\centering
\begin{tabular}{c|c|c|c|c|c|c|c|c|c|c||c|c}
\hline
\multicolumn{3}{c|}{} &  \multicolumn{2}{c|}{empty} & \multicolumn{2}{c|}{jumping} & \multicolumn{2}{c|}{standing} & \multicolumn{2}{c||}{walking} & \multicolumn{2}{c}{mean}\\\hline
area & rx. & antenna & Accuracy & F1 score & Accuracy & F1 score & Accuracy & F1 score& Accuracy & F1 score& Accuracy & F1 score\\\hline
\multirow{4}{*}{$\bm E_1$}& $1$ & $1$ &$100$ & $0.9967$ & $85.34$ & $0.9412$ & $82.67$ & $0.8575$ & $79.00$ & $0.8531$ &$87.00$& $0.9076$ \\
& $1$ & $4$ & $100$ & $0.9852$ & $98.00$ & $0.9484$ & $94.00$ & $0.9292$ & $85.67$ & $0.9212$ &$94.41$& $0.9460$\\
& $3$ & $1$ &$100$&$1.0000$ &$100$&$0.9891$ &$98.63$&$0.9905$ &$98.67$&$0.9928$ &$99.32$&$0.9931$ \\
& $3$ & $4$ & $100$ &$1.0000$ &$100$&$1.0000$ &$99.28$&$0.9960$ &$100$&$0.9980$ &$99.82$& $0.9977$\\\hline
\multirow{4}{*}{$\bm E_2$}& 1 & 1 &$100$&$0.9804$ &$78.00$&$0.7091$ &$62.00$&$0.6392$ &$73.00$&$0.8022$ &$78.25$&$0.7827$ \\
& $1$ & $4$ &$100$&$1.0000$ &$91.00$&$0.8778$ &$83.33$&$0.8460$ &$87.33$&$0.8927$ &$90.42$&$0.9041$\\
& $3$ & $1$ &$100$&$1.0000$ &$98.00$&$0.9018$ &$80.00$&$0.8775$ &$93.00$&$0.9884$ &$92.75$&$0.9417$\\
& $3$ & $4$ &$100$&$1.0000$ &$100$&$0.9800$ &$95.40$&$0.9853$ &$100$&$1.0000$ &$98.85$&$0.9085$
\\\hline
\multirow{4}{*}{$\bm E_3$}& 1 & 1 &$100$&$0.9967$ &$74.37$&$0.7348$ &$87.67$&$0.7799$ &$59.00$&$0.6933$ &$80.25$&$0.8012$ \\
& $1$ & $4$ &$100$&$0.9852$ &$95.00$&$0.8810$ &$92.00$&$0.9049$ &$75.67$&$0.8599$ &$90.67$&$0.9078$ \\
& $3$ & $1$ &$100$&$1.0000$ &$100$&$0.9747$ &$94.63$&$0.9649$ &$97.67$&$0.9827$ &$98.07$& $0.9806$\\
& $3$ & $4$ &$100$&$1.0000$ &$98.67$&$0.9852$ &$97.00$&$0.9848$ &$100$&$1.0000$ &$99.25$&$0.9925$ \\
\hline
\end{tabular}
\caption{Impact of macro- and macro- diversity on the accuracy of the predictions. \FW is trained in $E_1$ and is tested in $E_2$ and $E_3$.\vspace{-0.4cm} \label{tab:results}}
\end{table*}

\section{Performance Evaluation}\label{sec:perf_eva}

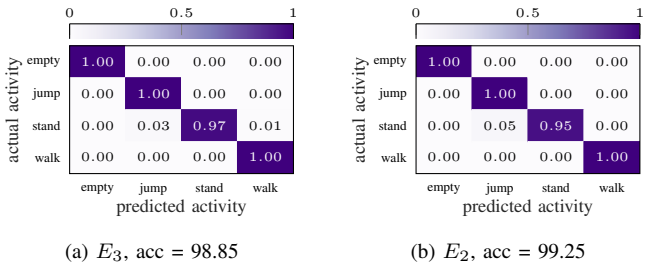
\begin{figure}[h]
    \begin{subfigure}[t]{0.48\columnwidth}
    \centering
    \setlength\fwidth{.7\columnwidth}
    \setlength\fheight{0.4\columnwidth}
    \begin{tikzpicture}
\pgfplotsset{every tick label/.append style={font=\tiny}}

\begin{axis}[
enlargelimits=false,
colorbar,
colormap/Purples,
width=\fwidth,
height=\fheight,
at={(0\fwidth,0\fheight)},
scale only axis,
tick align=inside,
xlabel={predicted activity},
xmin=-0.5,
xmax=3.5,
xtick style={draw=none},
xlabel style={font=\scriptsize\color{white!15!black}},
ylabel style={font=\scriptsize\color{white!15!black}},
ylabel={actual activity},
ymin=-0.5,
ymax=3.5,
xlabel shift=-5pt,
ylabel shift=-5pt,
xticklabels={empty, jump, stand, walk},
xtick={0,1,2,3},
ytick={0,1,2,3},
yticklabels={empty, jump, stand, walk},
ytick style={draw=none},
axis background/.style={fill=white},
colorbar horizontal,
colorbar style={
at={(0,1.05)},               
anchor=below south west,    
width=\pgfkeysvalueof{/pgfplots/parent axis width},
xtick={0, 0.5, 1},
xmin=0,
xmax=1,
axis x line*=top,
xticklabel shift=-1pt,
point meta min=0,
point meta max=1,
},
nodes near coords bottom/.style={
    scatter/position=absolute,
    close to zero/.style={at={(axis cs:\pgfkeysvalueof{/data point/x},\pgfkeysvalueof{/data point/y})}, xshift=0pt, yshift=-5pt, black, font=\tiny,
    /pgf/number format/.cd,
    fixed,
    fixed zerofill,
    precision=2,
    /tikz/.cd
    },
    big value/.style={at={(axis cs:\pgfkeysvalueof{/data point/x},\pgfkeysvalueof{/data point/y})}, xshift=0pt, yshift=-5pt, white, font=\tiny,
    /pgf/number format/.cd,
    fixed,
    fixed zerofill,
    precision=2,
    /tikz/.cd
    },
    every node near coord/.style={
      check for zero/.code={%
        \pgfmathfloatifflags{\pgfplotspointmeta}{0}{%
            \pgfkeys{/tikz/coordinate}%
        }{%
            \begingroup
            \pgfkeys{/pgf/fpu}%
            \pgfmathparse{\pgfplotspointmeta<#1}%
            \global\let\result=\pgfmathresult
            \endgroup
            %
            %
            \pgfmathfloatcreate{1}{1.0}{0}%
            \let\ONE=\pgfmathresult
            \ifx\result\ONE
                \pgfkeysalso{/pgfplots/close to zero}%
            \else
                \pgfkeysalso{/pgfplots/big value}%
            \fi
        }
      },
      check for zero, 
    },%
},%
nodes near coords bottom=0.5,
colorbar/width=2mm,
nodes near coords={\pgfmathprintnumber\pgfplotspointmeta},
]
\addplot [matrix plot,point meta=explicit]
 coordinates {
(0,0) [1] (0,1) [0.0001] (0,2) [.0001] (0,3) [.0001] 

(1,0) [0.0001] (1,1) [1] (1,2) [0.03] (1,3) [0.0001] 

(2,0) [0.0001] (2,1) [0.0001] (2,2) [0.97000] (2,3) [0.00001] 

(3,0) [0.0001] (3,1) [.0001] (3,2) [0.0133] (3,3) [1] 
};
\end{axis}
\end{tikzpicture}


%
%



%

    \caption{$E_3$, acc = 98.85}
    \label{fig:CM_80_3x4}
    \end{subfigure}
    \hfill
    \begin{subfigure}[t]{0.48\columnwidth}
    \centering
    \setlength\fwidth{.7\columnwidth}
    \setlength\fheight{0.4\columnwidth}
    \begin{tikzpicture}
\pgfplotsset{every tick label/.append style={font=\tiny}}

\begin{axis}[
enlargelimits=false,
colorbar,
colormap/Purples,
width=\fwidth,
height=\fheight,
at={(0\fwidth,0\fheight)},
scale only axis,
tick align=inside,
xlabel={predicted activity},
xmin=-0.5,
xmax=3.5,
xtick style={draw=none},
xlabel style={font=\scriptsize\color{white!15!black}},
ylabel style={font=\scriptsize\color{white!15!black}},
ylabel={actual activity},
ymin=-0.5,
ymax=3.5,
xlabel shift=-5pt,
ylabel shift=-5pt,
xticklabels={empty, jump, stand, walk},
xtick={0,1,2,3},
ytick={0,1,2,3},
yticklabels={empty, jump, stand, walk},
ytick style={draw=none},
axis background/.style={fill=white},
colorbar horizontal,
colorbar style={
at={(0,1.05)},               
anchor=below south west,    
width=\pgfkeysvalueof{/pgfplots/parent axis width},
xtick={0, 0.5, 1},
xmin=0,
xmax=1,
axis x line*=top,
xticklabel shift=-1pt,
point meta min=0,
point meta max=1,
},
nodes near coords bottom/.style={
    scatter/position=absolute,
    close to zero/.style={at={(axis cs:\pgfkeysvalueof{/data point/x},\pgfkeysvalueof{/data point/y})}, xshift=0pt, yshift=-5pt, black, font=\tiny,
    /pgf/number format/.cd,
    fixed,
    fixed zerofill,
    precision=2,
    /tikz/.cd
    },
    big value/.style={at={(axis cs:\pgfkeysvalueof{/data point/x},\pgfkeysvalueof{/data point/y})}, xshift=0pt, yshift=-5pt, white, font=\tiny,
    /pgf/number format/.cd,
    fixed,
    fixed zerofill,
    precision=2,
    /tikz/.cd
    },
    every node near coord/.style={
      check for zero/.code={%
        \pgfmathfloatifflags{\pgfplotspointmeta}{0}{%
            \pgfkeys{/tikz/coordinate}%
        }{%
            \begingroup
            \pgfkeys{/pgf/fpu}%
            \pgfmathparse{\pgfplotspointmeta<#1}%
            \global\let\result=\pgfmathresult
            \endgroup
            %
            %
            \pgfmathfloatcreate{1}{1.0}{0}%
            \let\ONE=\pgfmathresult
            \ifx\result\ONE
                \pgfkeysalso{/pgfplots/close to zero}%
            \else
                \pgfkeysalso{/pgfplots/big value}%
            \fi
        }
      },
      check for zero, 
    },%
},%
nodes near coords bottom=0.5,
colorbar/width=2mm,
nodes near coords={\pgfmathprintnumber\pgfplotspointmeta},
]
\addplot [matrix plot,point meta=explicit]
 coordinates {
(0,0) [1] (0,1) [0.0001] (0,2) [.0001] (0,3) [.0001] 

(1,0) [0.0001] (1,1) [1] (1,2) [0.046] (1,3) [0.0001]

(2,0) [0.0001] (2,1) [0.0001] (2,2) [0.954] (2,3) [0.0001] 

(3,0) [0.0001] (3,1) [.0001] (3,2) [0.0001] (3,3) [1] 
};
\end{axis}
\end{tikzpicture}


%
%



%

    \caption{$E_2$, acc = 99.25}
    \label{fig:CM_80_3x4_A3}
    \end{subfigure}
    \caption{The performance of \FW in generalizing to $E_2$ and $E_3$ environment, with 3 receivers, 4 antennas, 80 MHz.\vspace{-0.3cm}}
        \label{fig:overall_performance}
\end{figure}

In this section, we show the end-to-end performance evaluation of \FW. Figure \ref{fig:overall_performance} shows the overall performance of \FW  in terms of accuracy of predictions with 3 receivers, each with 4 antennas, at 80 MHz channels when tested in environment $E_2$ and $E_3$. It can be seen that \FW not only learned very well from the source dataset but successfully generalized to the target environments, as it is able to achieve accuracy close to 100\% in unseen environments by taking advantage of all three components of macro/micro diversity and subcarrier resolution, in addition to implementing \gls{prnet} with modified embedding learning.

\subsubsection{Impact of Macro- and Micro-diversity}

Table \ref{tab:results} shows the impact of macro- and micro- diversity on the accuracy of the predictions. It can be seen that increasing the number of antennas from 1 to 4 increases the accuracy by 12\% and 10\% in environments $E_2$ and $E_3$, respectively. In addition, increasing the number of receivers from 1 to 3 improves the accuracy by 14\% and 18\% in environments $E_2$ and $E_3$, respectively. By looking closely at these results, we can see that in general, micro-diversity (number of antennas) improves the accuracy of in-place activities, such as jump and standing, while macro-diversity improves the accuracy of walking. Interestingly, macro-diversity helps discriminate walking from standing and jumping and micro-diversity increases the precision of differentiating jumping from standing.

\subsubsection{Impact of Subcarrier Resolution}

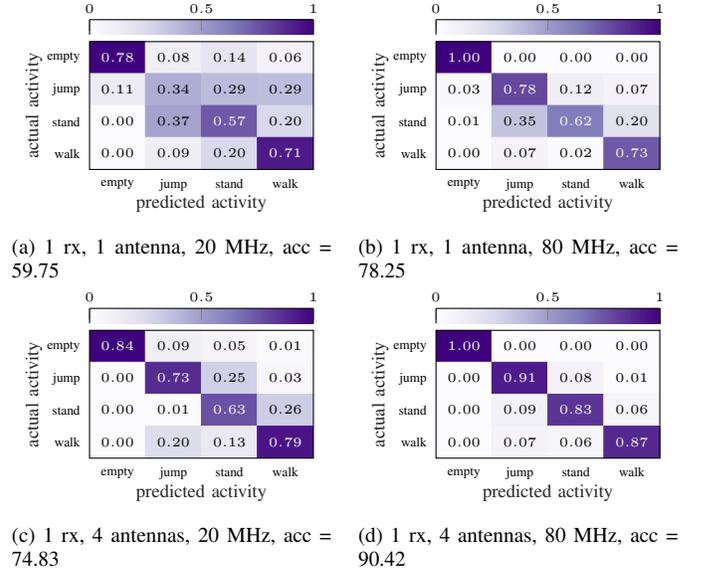
\begin{figure}[!ht]
 \begin{subfigure}[t]{0.48\columnwidth}
    \centering
    \setlength\fwidth{.7\columnwidth}
    \setlength\fheight{0.4\columnwidth}
    \begin{tikzpicture}
\pgfplotsset{every tick label/.append style={font=\tiny}}

\begin{axis}[
enlargelimits=false,
colorbar,
colormap/Purples,
width=\fwidth,
height=\fheight,
at={(0\fwidth,0\fheight)},
scale only axis,
tick align=inside,
xlabel={predicted activity},
xmin=-0.5,
xmax=3.5,
xtick style={draw=none},
xlabel style={font=\scriptsize\color{white!15!black}},
ylabel style={font=\scriptsize\color{white!15!black}},
ylabel={actual activity},
ymin=-0.5,
ymax=3.5,
xlabel shift=-5pt,
ylabel shift=-5pt,
xticklabels={empty, jump, stand, walk},
xtick={0,1,2,3},
ytick={0,1,2,3},
yticklabels={empty, jump, stand, walk},
ytick style={draw=none},
axis background/.style={fill=white},
colorbar horizontal,
colorbar style={
at={(0,1.05)},               
anchor=below south west,    
width=\pgfkeysvalueof{/pgfplots/parent axis width},
xtick={0, 0.5, 1},
xmin=0,
xmax=1,
axis x line*=top,
xticklabel shift=-1pt,
point meta min=0,
point meta max=1,
},
nodes near coords bottom/.style={
    scatter/position=absolute,
    close to zero/.style={at={(axis cs:\pgfkeysvalueof{/data point/x},\pgfkeysvalueof{/data point/y})}, xshift=0pt, yshift=-5pt, black, font=\tiny,
    /pgf/number format/.cd,
    fixed,
    fixed zerofill,
    precision=2,
    /tikz/.cd
    },
    big value/.style={at={(axis cs:\pgfkeysvalueof{/data point/x},\pgfkeysvalueof{/data point/y})}, xshift=0pt, yshift=-5pt, white, font=\tiny,
    /pgf/number format/.cd,
    fixed,
    fixed zerofill,
    precision=2,
    /tikz/.cd
    },
    every node near coord/.style={
      check for zero/.code={%
        \pgfmathfloatifflags{\pgfplotspointmeta}{0}{%
            \pgfkeys{/tikz/coordinate}%
        }{%
            \begingroup
            \pgfkeys{/pgf/fpu}%
            \pgfmathparse{\pgfplotspointmeta<#1}%
            \global\let\result=\pgfmathresult
            \endgroup
            %
            %
            \pgfmathfloatcreate{1}{1.0}{0}%
            \let\ONE=\pgfmathresult
            \ifx\result\ONE
                \pgfkeysalso{/pgfplots/close to zero}%
            \else
                \pgfkeysalso{/pgfplots/big value}%
            \fi
        }
      },
      check for zero, 
    },%
},%
nodes near coords bottom=0.5,
colorbar/width=2mm,
nodes near coords={\pgfmathprintnumber\pgfplotspointmeta},
]
\addplot [matrix plot,point meta=explicit]
 coordinates {
(0,0) [0.7767] (0,1) [0.1056] (0,2) [.0001] (0,3) [.0001] 

(1,0) [0.0833] (1,1) [0.3367] (1,2) [0.3667] (1,3) [0.0933] 

(2,0) [0.1400] (2,1) [0.2933] (2,2) [0.5667] (2,3) [0.1967] 

(3,0) [0.0600] (3,1) [0.2867] (3,2) [0.1967] (3,3) [0.7133] 
};
\end{axis}
\end{tikzpicture}


%
%



%

    \caption{1 rx, 1 antenna, 20 MHz, acc = 59.75}
    \end{subfigure}
        \hfill
    \centering
    \begin{subfigure}[t]{0.48\columnwidth}
    \centering
    \setlength\fwidth{.7\columnwidth}
    \setlength\fheight{0.4\columnwidth}
    \begin{tikzpicture}
\pgfplotsset{every tick label/.append style={font=\tiny}}

\begin{axis}[
enlargelimits=false,
colorbar,
colormap/Purples,
width=\fwidth,
height=\fheight,
at={(0\fwidth,0\fheight)},
scale only axis,
tick align=inside,
xlabel={predicted activity},
xmin=-0.5,
xmax=3.5,
xtick style={draw=none},
xlabel style={font=\scriptsize\color{white!15!black}},
ylabel style={font=\scriptsize\color{white!15!black}},
ylabel={actual activity},
ymin=-0.5,
ymax=3.5,
xlabel shift=-5pt,
ylabel shift=-5pt,
xticklabels={empty, jump, stand, walk},
xtick={0,1,2,3},
ytick={0,1,2,3},
yticklabels={empty, jump, stand, walk},
ytick style={draw=none},
axis background/.style={fill=white},
colorbar horizontal,
colorbar style={
at={(0,1.05)},               
anchor=below south west,    
width=\pgfkeysvalueof{/pgfplots/parent axis width},
xtick={0, 0.5, 1},
xmin=0,
xmax=1,
axis x line*=top,
xticklabel shift=-1pt,
point meta min=0,
point meta max=1,
},
nodes near coords bottom/.style={
    scatter/position=absolute,
    close to zero/.style={at={(axis cs:\pgfkeysvalueof{/data point/x},\pgfkeysvalueof{/data point/y})}, xshift=0pt, yshift=-5pt, black, font=\tiny,
    /pgf/number format/.cd,
    fixed,
    fixed zerofill,
    precision=2,
    /tikz/.cd
    },
    big value/.style={at={(axis cs:\pgfkeysvalueof{/data point/x},\pgfkeysvalueof{/data point/y})}, xshift=0pt, yshift=-5pt, white, font=\tiny,
    /pgf/number format/.cd,
    fixed,
    fixed zerofill,
    precision=2,
    /tikz/.cd
    },
    every node near coord/.style={
      check for zero/.code={%
        \pgfmathfloatifflags{\pgfplotspointmeta}{0}{%
            \pgfkeys{/tikz/coordinate}%
        }{%
            \begingroup
            \pgfkeys{/pgf/fpu}%
            \pgfmathparse{\pgfplotspointmeta<#1}%
            \global\let\result=\pgfmathresult
            \endgroup
            %
            %
            \pgfmathfloatcreate{1}{1.0}{0}%
            \let\ONE=\pgfmathresult
            \ifx\result\ONE
                \pgfkeysalso{/pgfplots/close to zero}%
            \else
                \pgfkeysalso{/pgfplots/big value}%
            \fi
        }
      },
      check for zero, 
    },%
},%
nodes near coords bottom=0.5,
colorbar/width=2mm,
nodes near coords={\pgfmathprintnumber\pgfplotspointmeta},
]
\addplot [matrix plot,point meta=explicit]
 coordinates {
(0,0) [1] (0,1) [0.0301] (0,2) [0.0101] (0,3) [.0001] 

(1,0) [0.0001] (1,1) [0.78] (1,2) [0.35] (1,3) [0.07] 

(2,0) [0.0001] (2,1) [0.12] (2,2) [0.62] (2,3) [0.02] 

(3,0) [0.0001] (3,1) [0.07] (3,2) [0.2] (3,3) [0.73] 
};
\end{axis}
\end{tikzpicture}


%
%



%

    \caption{1 rx, 1 antenna, 80 MHz, acc = 78.25}
    \end{subfigure}
        \hfill
     \begin{subfigure}[t]{0.48\columnwidth}
    \centering
    \setlength\fwidth{.7\columnwidth}
    \setlength\fheight{0.4\columnwidth}
    \begin{tikzpicture}
\pgfplotsset{every tick label/.append style={font=\tiny}}

\begin{axis}[
enlargelimits=false,
colorbar,
colormap/Purples,
width=\fwidth,
height=\fheight,
at={(0\fwidth,0\fheight)},
scale only axis,
tick align=inside,
xlabel={predicted activity},
xmin=-0.5,
xmax=3.5,
xtick style={draw=none},
xlabel style={font=\scriptsize\color{white!15!black}},
ylabel style={font=\scriptsize\color{white!15!black}},
ylabel={actual activity},
ymin=-0.5,
ymax=3.5,
xlabel shift=-5pt,
ylabel shift=-5pt,
xticklabels={empty, jump, stand, walk},
xtick={0,1,2,3},
ytick={0,1,2,3},
yticklabels={empty, jump, stand, walk},
ytick style={draw=none},
axis background/.style={fill=white},
colorbar horizontal,
colorbar style={
at={(0,1.05)},               
anchor=below south west,    
width=\pgfkeysvalueof{/pgfplots/parent axis width},
xtick={0, 0.5, 1},
xmin=0,
xmax=1,
axis x line*=top,
xticklabel shift=-1pt,
point meta min=0,
point meta max=1,
},
nodes near coords bottom/.style={
    scatter/position=absolute,
    close to zero/.style={at={(axis cs:\pgfkeysvalueof{/data point/x},\pgfkeysvalueof{/data point/y})}, xshift=0pt, yshift=-5pt, black, font=\tiny,
    /pgf/number format/.cd,
    fixed,
    fixed zerofill,
    precision=2,
    /tikz/.cd
    },
    big value/.style={at={(axis cs:\pgfkeysvalueof{/data point/x},\pgfkeysvalueof{/data point/y})}, xshift=0pt, yshift=-5pt, white, font=\tiny,
    /pgf/number format/.cd,
    fixed,
    fixed zerofill,
    precision=2,
    /tikz/.cd
    },
    every node near coord/.style={
      check for zero/.code={%
        \pgfmathfloatifflags{\pgfplotspointmeta}{0}{%
            \pgfkeys{/tikz/coordinate}%
        }{%
            \begingroup
            \pgfkeys{/pgf/fpu}%
            \pgfmathparse{\pgfplotspointmeta<#1}%
            \global\let\result=\pgfmathresult
            \endgroup
            %
            %
            \pgfmathfloatcreate{1}{1.0}{0}%
            \let\ONE=\pgfmathresult
            \ifx\result\ONE
                \pgfkeysalso{/pgfplots/close to zero}%
            \else
                \pgfkeysalso{/pgfplots/big value}%
            \fi
        }
      },
      check for zero, 
    },%
},%
nodes near coords bottom=0.5,
colorbar/width=2mm,
nodes near coords={\pgfmathprintnumber\pgfplotspointmeta},
]
\addplot [matrix plot,point meta=explicit]
 coordinates {
(0,0) [0.84330] (0,1) [0.0001] (0,2) [0.0001] (0,3) [.0001] 

(1,0) [0.09] (1,1) [0.7267] (1,2) [0.011] (1,3) [0.1967] 

(2,0) [0.05] (2,1) [0.2467] (2,2) [0.6333] (2,3) [0.133] 

(3,0) [0.0133] (3,1) [0.0267] (3,2) [0.2567] (3,3) [0.79] 
};
\end{axis}
\end{tikzpicture}


%
%



%

    \caption{1 rx, 4 antennas, 20 MHz, acc = 74.83}
    \end{subfigure}
        \hfill
            \begin{subfigure}[t]{0.48\columnwidth}
    \centering
    \setlength\fwidth{.7\columnwidth}
    \setlength\fheight{0.4\columnwidth}
    \begin{tikzpicture}
\pgfplotsset{every tick label/.append style={font=\tiny}}

\begin{axis}[
enlargelimits=false,
colorbar,
colormap/Purples,
width=\fwidth,
height=\fheight,
at={(0\fwidth,0\fheight)},
scale only axis,
tick align=inside,
xlabel={predicted activity},
xmin=-0.5,
xmax=3.5,
xtick style={draw=none},
xlabel style={font=\scriptsize\color{white!15!black}},
ylabel style={font=\scriptsize\color{white!15!black}},
ylabel={actual activity},
ymin=-0.5,
ymax=3.5,
xlabel shift=-5pt,
ylabel shift=-5pt,
xticklabels={empty, jump, stand, walk},
xtick={0,1,2,3},
ytick={0,1,2,3},
yticklabels={empty, jump, stand, walk},
ytick style={draw=none},
axis background/.style={fill=white},
colorbar horizontal,
colorbar style={
at={(0,1.05)},               
anchor=below south west,    
width=\pgfkeysvalueof{/pgfplots/parent axis width},
xtick={0, 0.5, 1},
xmin=0,
xmax=1,
axis x line*=top,
xticklabel shift=-1pt,
point meta min=0,
point meta max=1,
},
nodes near coords bottom/.style={
    scatter/position=absolute,
    close to zero/.style={at={(axis cs:\pgfkeysvalueof{/data point/x},\pgfkeysvalueof{/data point/y})}, xshift=0pt, yshift=-5pt, black, font=\tiny,
    /pgf/number format/.cd,
    fixed,
    fixed zerofill,
    precision=2,
    /tikz/.cd
    },
    big value/.style={at={(axis cs:\pgfkeysvalueof{/data point/x},\pgfkeysvalueof{/data point/y})}, xshift=0pt, yshift=-5pt, white, font=\tiny,
    /pgf/number format/.cd,
    fixed,
    fixed zerofill,
    precision=2,
    /tikz/.cd
    },
    every node near coord/.style={
      check for zero/.code={%
        \pgfmathfloatifflags{\pgfplotspointmeta}{0}{%
            \pgfkeys{/tikz/coordinate}%
        }{%
            \begingroup
            \pgfkeys{/pgf/fpu}%
            \pgfmathparse{\pgfplotspointmeta<#1}%
            \global\let\result=\pgfmathresult
            \endgroup
            %
            %
            \pgfmathfloatcreate{1}{1.0}{0}%
            \let\ONE=\pgfmathresult
            \ifx\result\ONE
                \pgfkeysalso{/pgfplots/close to zero}%
            \else
                \pgfkeysalso{/pgfplots/big value}%
            \fi
        }
      },
      check for zero, 
    },%
},%
nodes near coords bottom=0.5,
colorbar/width=2mm,
nodes near coords={\pgfmathprintnumber\pgfplotspointmeta},
]
\addplot [matrix plot,point meta=explicit]
 coordinates {
(0,0) [1] (0,1) [0.0001] (0,2) [.0001] (0,3) [.0001] 

(1,0) [0.0001] (1,1) [0.91] (1,2) [0.0933] (1,3) [0.07] 

(2,0) [0.0001] (2,1) [0.08] (2,2) [0.8333] (2,3) [0.0567] 

(3,0) [0.0001] (3,1) [0.01] (3,2) [0.0567] (3,3) [0.8733] 
};
\end{axis}
\end{tikzpicture}


%
%



%

    \caption{1 rx, 4 antennas, 80 MHz, acc = 90.42}
    \end{subfigure}
    \caption{Impact of frequency resolution and diversity, with one receiver. $E_2$ is the target environment. \vspace{-0.3cm} }
    \label{fig:resolution_1rx}
\end{figure}

\begin{figure}[!ht]
      \begin{subfigure}[t]{0.48\columnwidth}
    \centering
    \setlength\fwidth{.7\columnwidth}
    \setlength\fheight{0.4\columnwidth}
    \begin{tikzpicture}
\pgfplotsset{every tick label/.append style={font=\tiny}}

\begin{axis}[
enlargelimits=false,
colorbar,
colormap/Purples,
width=\fwidth,
height=\fheight,
at={(0\fwidth,0\fheight)},
scale only axis,
tick align=inside,
xlabel={predicted activity},
xmin=-0.5,
xmax=3.5,
xtick style={draw=none},
xlabel style={font=\scriptsize\color{white!15!black}},
ylabel style={font=\scriptsize\color{white!15!black}},
ylabel={actual activity},
ymin=-0.5,
ymax=3.5,
xlabel shift=-5pt,
ylabel shift=-5pt,
xticklabels={empty, jump, stand, walk},
xtick={0,1,2,3},
ytick={0,1,2,3},
yticklabels={empty, jump, stand, walk},
ytick style={draw=none},
axis background/.style={fill=white},
colorbar horizontal,
colorbar style={
at={(0,1.05)},               
anchor=below south west,    
width=\pgfkeysvalueof{/pgfplots/parent axis width},
xtick={0, 0.5, 1},
xmin=0,
xmax=1,
axis x line*=top,
xticklabel shift=-1pt,
point meta min=0,
point meta max=1,
},
nodes near coords bottom/.style={
    scatter/position=absolute,
    close to zero/.style={at={(axis cs:\pgfkeysvalueof{/data point/x},\pgfkeysvalueof{/data point/y})}, xshift=0pt, yshift=-5pt, black, font=\tiny,
    /pgf/number format/.cd,
    fixed,
    fixed zerofill,
    precision=2,
    /tikz/.cd
    },
    big value/.style={at={(axis cs:\pgfkeysvalueof{/data point/x},\pgfkeysvalueof{/data point/y})}, xshift=0pt, yshift=-5pt, white, font=\tiny,
    /pgf/number format/.cd,
    fixed,
    fixed zerofill,
    precision=2,
    /tikz/.cd
    },
    every node near coord/.style={
      check for zero/.code={%
        \pgfmathfloatifflags{\pgfplotspointmeta}{0}{%
            \pgfkeys{/tikz/coordinate}%
        }{%
            \begingroup
            \pgfkeys{/pgf/fpu}%
            \pgfmathparse{\pgfplotspointmeta<#1}%
            \global\let\result=\pgfmathresult
            \endgroup
            %
            %
            \pgfmathfloatcreate{1}{1.0}{0}%
            \let\ONE=\pgfmathresult
            \ifx\result\ONE
                \pgfkeysalso{/pgfplots/close to zero}%
            \else
                \pgfkeysalso{/pgfplots/big value}%
            \fi
        }
      },
      check for zero, 
    },%
},%
nodes near coords bottom=0.5,
colorbar/width=2mm,
nodes near coords={\pgfmathprintnumber\pgfplotspointmeta},
]
\addplot [matrix plot,point meta=explicit]
 coordinates {
(0,0) [1] (0,1) [0.0001] (0,2) [0.0033] (0,3) [.0001] 

(1,0) [0.0001] (1,1) [0.4967] (1,2) [0.333] (1,3) [0.07] 

(2,0) [0.0001] (2,1) [0.2333] (2,2) [0.6333] (2,3) [0.03] 

(3,0) [0.0001] (3,1) [0.27] (3,2) [0.0001] (3,3) [0.9] 
};
\end{axis}
\end{tikzpicture}


%
%



%

    \caption{3 rx, 1 antenna, 20 MHz, acc = 76.5}
    \end{subfigure}
    \hfill
    \begin{subfigure}[t]{0.48\columnwidth}
    \centering
    \setlength\fwidth{.7\columnwidth}
    \setlength\fheight{0.4\columnwidth}
    \begin{tikzpicture}
\pgfplotsset{every tick label/.append style={font=\tiny}}

\begin{axis}[
enlargelimits=false,
colorbar,
colormap/Purples,
width=\fwidth,
height=\fheight,
at={(0\fwidth,0\fheight)},
scale only axis,
tick align=inside,
xlabel={predicted activity},
xmin=-0.5,
xmax=3.5,
xtick style={draw=none},
xlabel style={font=\scriptsize\color{white!15!black}},
ylabel style={font=\scriptsize\color{white!15!black}},
ylabel={actual activity},
ymin=-0.5,
ymax=3.5,
xlabel shift=-5pt,
ylabel shift=-5pt,
xticklabels={empty, jump, stand, walk},
xtick={0,1,2,3},
ytick={0,1,2,3},
yticklabels={empty, jump, stand, walk},
ytick style={draw=none},
axis background/.style={fill=white},
colorbar horizontal,
colorbar style={
at={(0,1.05)},               
anchor=below south west,    
width=\pgfkeysvalueof{/pgfplots/parent axis width},
xtick={0, 0.5, 1},
xmin=0,
xmax=1,
axis x line*=top,
xticklabel shift=-1pt,
point meta min=0,
point meta max=1,
},
nodes near coords bottom/.style={
    scatter/position=absolute,
    close to zero/.style={at={(axis cs:\pgfkeysvalueof{/data point/x},\pgfkeysvalueof{/data point/y})}, xshift=0pt, yshift=-5pt, black, font=\tiny,
    /pgf/number format/.cd,
    fixed,
    fixed zerofill,
    precision=2,
    /tikz/.cd
    },
    big value/.style={at={(axis cs:\pgfkeysvalueof{/data point/x},\pgfkeysvalueof{/data point/y})}, xshift=0pt, yshift=-5pt, white, font=\tiny,
    /pgf/number format/.cd,
    fixed,
    fixed zerofill,
    precision=2,
    /tikz/.cd
    },
    every node near coord/.style={
      check for zero/.code={%
        \pgfmathfloatifflags{\pgfplotspointmeta}{0}{%
            \pgfkeys{/tikz/coordinate}%
        }{%
            \begingroup
            \pgfkeys{/pgf/fpu}%
            \pgfmathparse{\pgfplotspointmeta<#1}%
            \global\let\result=\pgfmathresult
            \endgroup
            %
            %
            \pgfmathfloatcreate{1}{1.0}{0}%
            \let\ONE=\pgfmathresult
            \ifx\result\ONE
                \pgfkeysalso{/pgfplots/close to zero}%
            \else
                \pgfkeysalso{/pgfplots/big value}%
            \fi
        }
      },
      check for zero, 
    },%
},%
nodes near coords bottom=0.5,
colorbar/width=2mm,
nodes near coords={\pgfmathprintnumber\pgfplotspointmeta},
]
\addplot [matrix plot,point meta=explicit]
 coordinates {
(0,0) [1] (0,1) [0.0001] (0,2) [.0001] (0,3) [.0001] 

(1,0) [0.0001] (1,1) [0.98] (1,2) [0.1867] (1,3) [0.0067] 

(2,0) [0.0001] (2,1) [0.02] (2,2) [0.8000] (2,3) [0.0033] 

(3,0) [0.0001] (3,1) [.0001] (3,2) [0.0133] (3,3) [0.93] 
};
\end{axis}
\end{tikzpicture}


%
%



%

    \caption{3 rx, 1 antenna, 80 MHz, acc = 92.75}
    \end{subfigure}
        \setlength\abovecaptionskip{0.2cm}
    \hfill
    \begin{subfigure}[t]{0.48\columnwidth}
    \centering
    \setlength\fwidth{.7\columnwidth}
    \setlength\fheight{0.4\columnwidth}
    \begin{tikzpicture}
\pgfplotsset{every tick label/.append style={font=\tiny}}

\begin{axis}[
enlargelimits=false,
colorbar,
colormap/Purples,
width=\fwidth,
height=\fheight,
at={(0\fwidth,0\fheight)},
scale only axis,
tick align=inside,
xlabel={predicted activity},
xmin=-0.5,
xmax=3.5,
xtick style={draw=none},
xlabel style={font=\scriptsize\color{white!15!black}},
ylabel style={font=\scriptsize\color{white!15!black}},
ylabel={actual activity},
ymin=-0.5,
ymax=3.5,
xlabel shift=-5pt,
ylabel shift=-5pt,
xticklabels={empty, jump, stand, walk},
xtick={0,1,2,3},
ytick={0,1,2,3},
yticklabels={empty, jump, stand, walk},
ytick style={draw=none},
axis background/.style={fill=white},
colorbar horizontal,
colorbar style={
at={(0,1.05)},               
anchor=below south west,    
width=\pgfkeysvalueof{/pgfplots/parent axis width},
xtick={0, 0.5, 1},
xmin=0,
xmax=1,
axis x line*=top,
xticklabel shift=-1pt,
point meta min=0,
point meta max=1,
},
nodes near coords bottom/.style={
    scatter/position=absolute,
    close to zero/.style={at={(axis cs:\pgfkeysvalueof{/data point/x},\pgfkeysvalueof{/data point/y})}, xshift=0pt, yshift=-5pt, black, font=\tiny,
    /pgf/number format/.cd,
    fixed,
    fixed zerofill,
    precision=2,
    /tikz/.cd
    },
    big value/.style={at={(axis cs:\pgfkeysvalueof{/data point/x},\pgfkeysvalueof{/data point/y})}, xshift=0pt, yshift=-5pt, white, font=\tiny,
    /pgf/number format/.cd,
    fixed,
    fixed zerofill,
    precision=2,
    /tikz/.cd
    },
    every node near coord/.style={
      check for zero/.code={%
        \pgfmathfloatifflags{\pgfplotspointmeta}{0}{%
            \pgfkeys{/tikz/coordinate}%
        }{%
            \begingroup
            \pgfkeys{/pgf/fpu}%
            \pgfmathparse{\pgfplotspointmeta<#1}%
            \global\let\result=\pgfmathresult
            \endgroup
            %
            %
            \pgfmathfloatcreate{1}{1.0}{0}%
            \let\ONE=\pgfmathresult
            \ifx\result\ONE
                \pgfkeysalso{/pgfplots/close to zero}%
            \else
                \pgfkeysalso{/pgfplots/big value}%
            \fi
        }
      },
      check for zero, 
    },%
},%
nodes near coords bottom=0.5,
colorbar/width=2mm,
nodes near coords={\pgfmathprintnumber\pgfplotspointmeta},
]
\addplot [matrix plot,point meta=explicit]
 coordinates {
(0,0) [.9067] (0,1) [0.0001] (0,2) [0.0000] (0,3) [.0001] 

(1,0) [0.0001] (1,1) [0.8067] (1,2) [0.2233] (1,3) [0.02] 

(2,0) [0.0001] (2,1) [0.12] (2,2) [0.6767] (2,3) [0.01] 

(3,0) [0.0093] (3,1) [0.0733] (3,2) [0.1] (3,3) [0.96] 
};
\end{axis}
\end{tikzpicture}


%
%



%

    \caption{3 rx, 4 antennas, 20 MHz, acc = 84.25}
    \end{subfigure}
    \hfill
    \begin{subfigure}[t]{0.48\columnwidth}
    \centering
    \setlength\fwidth{.7\columnwidth}
    \setlength\fheight{0.4\columnwidth}
    \begin{tikzpicture}
\pgfplotsset{every tick label/.append style={font=\tiny}}

\begin{axis}[
enlargelimits=false,
colorbar,
colormap/Purples,
width=\fwidth,
height=\fheight,
at={(0\fwidth,0\fheight)},
scale only axis,
tick align=inside,
xlabel={predicted activity},
xmin=-0.5,
xmax=3.5,
xtick style={draw=none},
xlabel style={font=\scriptsize\color{white!15!black}},
ylabel style={font=\scriptsize\color{white!15!black}},
ylabel={actual activity},
ymin=-0.5,
ymax=3.5,
xlabel shift=-5pt,
ylabel shift=-5pt,
xticklabels={empty, jump, stand, walk},
xtick={0,1,2,3},
ytick={0,1,2,3},
yticklabels={empty, jump, stand, walk},
ytick style={draw=none},
axis background/.style={fill=white},
colorbar horizontal,
colorbar style={
at={(0,1.05)},               
anchor=below south west,    
width=\pgfkeysvalueof{/pgfplots/parent axis width},
xtick={0, 0.5, 1},
xmin=0,
xmax=1,
axis x line*=top,
xticklabel shift=-1pt,
point meta min=0,
point meta max=1,
},
nodes near coords bottom/.style={
    scatter/position=absolute,
    close to zero/.style={at={(axis cs:\pgfkeysvalueof{/data point/x},\pgfkeysvalueof{/data point/y})}, xshift=0pt, yshift=-5pt, black, font=\tiny,
    /pgf/number format/.cd,
    fixed,
    fixed zerofill,
    precision=2,
    /tikz/.cd
    },
    big value/.style={at={(axis cs:\pgfkeysvalueof{/data point/x},\pgfkeysvalueof{/data point/y})}, xshift=0pt, yshift=-5pt, white, font=\tiny,
    /pgf/number format/.cd,
    fixed,
    fixed zerofill,
    precision=2,
    /tikz/.cd
    },
    every node near coord/.style={
      check for zero/.code={%
        \pgfmathfloatifflags{\pgfplotspointmeta}{0}{%
            \pgfkeys{/tikz/coordinate}%
        }{%
            \begingroup
            \pgfkeys{/pgf/fpu}%
            \pgfmathparse{\pgfplotspointmeta<#1}%
            \global\let\result=\pgfmathresult
            \endgroup
            %
            %
            \pgfmathfloatcreate{1}{1.0}{0}%
            \let\ONE=\pgfmathresult
            \ifx\result\ONE
                \pgfkeysalso{/pgfplots/close to zero}%
            \else
                \pgfkeysalso{/pgfplots/big value}%
            \fi
        }
      },
      check for zero, 
    },%
},%
nodes near coords bottom=0.5,
colorbar/width=2mm,
nodes near coords={\pgfmathprintnumber\pgfplotspointmeta},
]
\addplot [matrix plot,point meta=explicit]
 coordinates {
(0,0) [1] (0,1) [0.0001] (0,2) [.0001] (0,3) [.0001] 

(1,0) [0.0001] (1,1) [1] (1,2) [0.046] (1,3) [0.0001]

(2,0) [0.0001] (2,1) [0.0001] (2,2) [0.954] (2,3) [0.0001] 

(3,0) [0.0001] (3,1) [.0001] (3,2) [0.0001] (3,3) [1] 
};
\end{axis}
\end{tikzpicture}


%
%



%

    \caption{3 rx, 4 antennas, 80 MHz, acc = 98.85}
    \end{subfigure}
    \caption{Impact of frequency resolution and diversity, with one receiver. $E_2$ is the target environment.\vspace{-0.3cm}}
        \label{fig:resolution_3rx}
\end{figure}
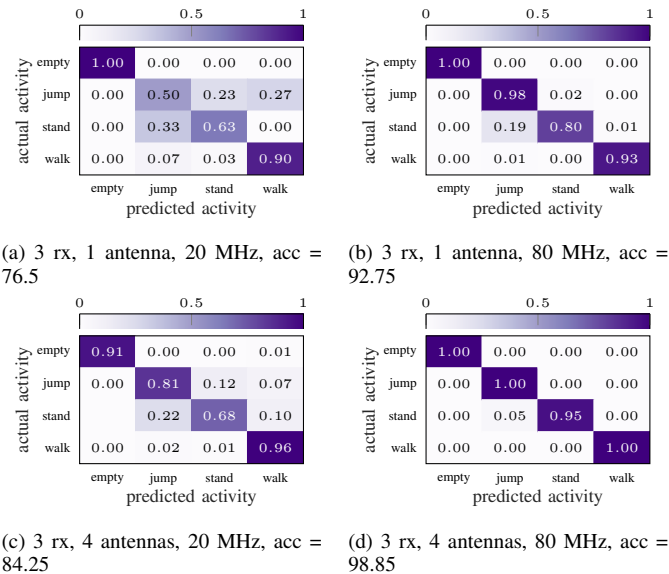

Figures \ref{fig:resolution_1rx} and \ref{fig:resolution_3rx} show the impact of subcarrier resolution and the number of antennas by comparing the confusion matrices of \FW in 20 MHz and 80 MHz channels, respectively with 1 receiver and 3 receivers. It can be seen that at 20 MHz with 1 receiver the accuracy diminishes by close to 20\% in the worst case. The number of receivers improves the performance, but we still notice that higher resolution implies 6\% better accuracy. In general, differentiating standing from jumping and walking is a challenging task, since our human subjects were allowed to make small body movements like moving arms and head (as it is naturally). Based on this result, we conclude that higher frequency can be beneficial for detecting activities with micro-movements like respiratory detection \cite{wifiRespiratoryWide}.

\subsubsection{Comparison of CNN vs. \FW}

\begin{figure}[h!]
    \centering
    \includegraphics[width=1\columnwidth,trim={0cm 1cm 0 2cm},clip]{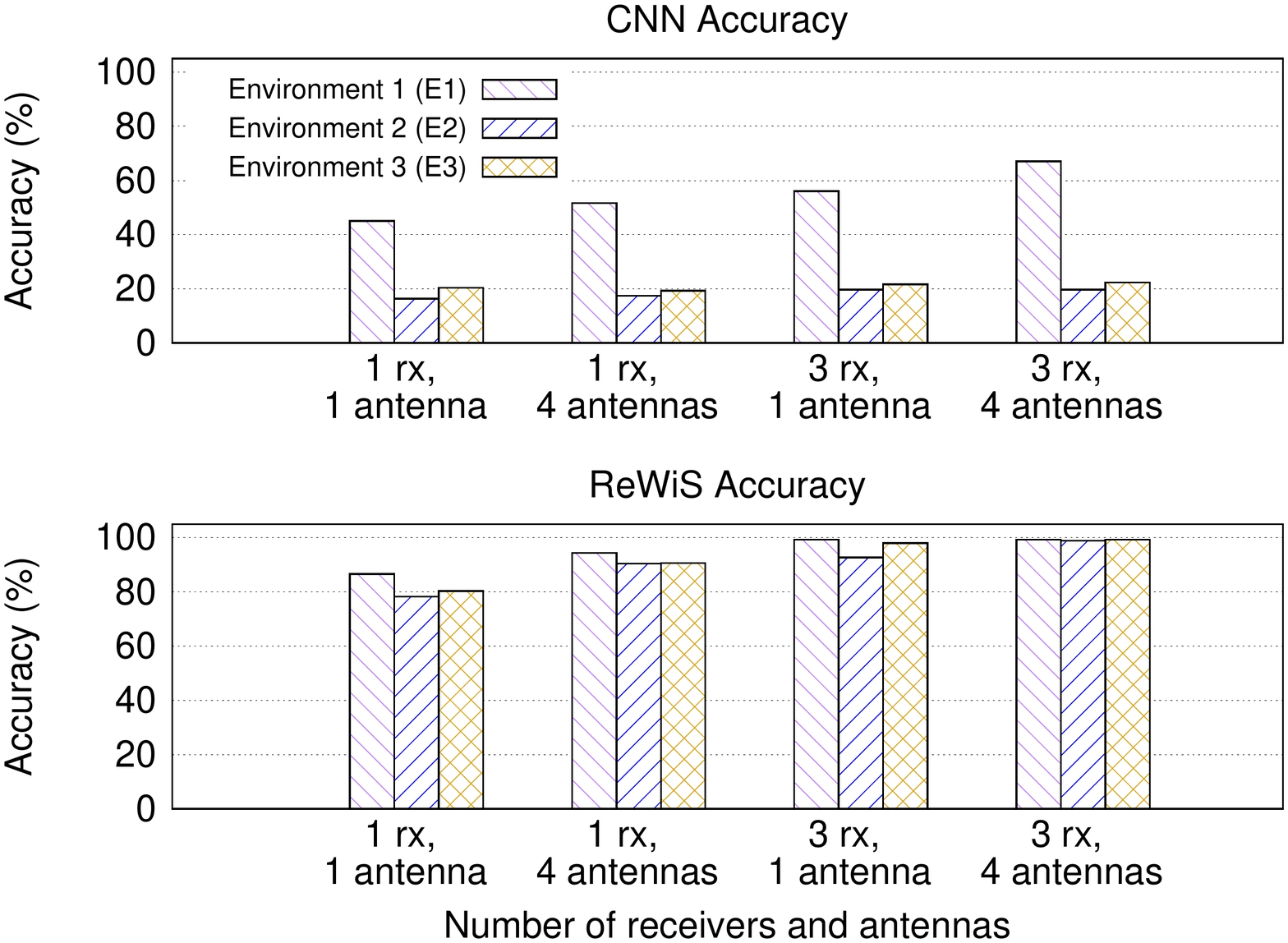}
    \caption{CNN vs \FW accuracy performance.\vspace{-0.2cm}}
        \label{fig:cnn_vs_fw}
\end{figure}

Figure \ref{fig:cnn_vs_fw} compares the performance of \FW learning framework with a baseline \gls{cnn} classifier \cite{muaaz2021wi}. The classifier utilizes a \gls{cnn} for feature extraction with the same structure as the one \FW uses for embedding function training, explained in Section \ref{sec:learning}. Further, a 3-layer fully connected network is used for classification. Overall, the total number of parameters of the baseline classifier is comparable with the \FW learning framework. We notice that the baseline model is able to learn from the \gls{csi} data collected in the source environment $E_1$, however, is not able to make accurate predictions in target environments. This happens since the baseline classifier's goal is to classify the \gls{csi} measurement. However, the \FW learning framework tends to learn \textit{how} to learn and achieves 35\% better accuracy on average.

\subsubsection{Impact of windowing}

\begin{figure}[h!]
    \centering
    \includegraphics[width=.9\columnwidth,trim={0cm 4cm 0 3cm},clip]{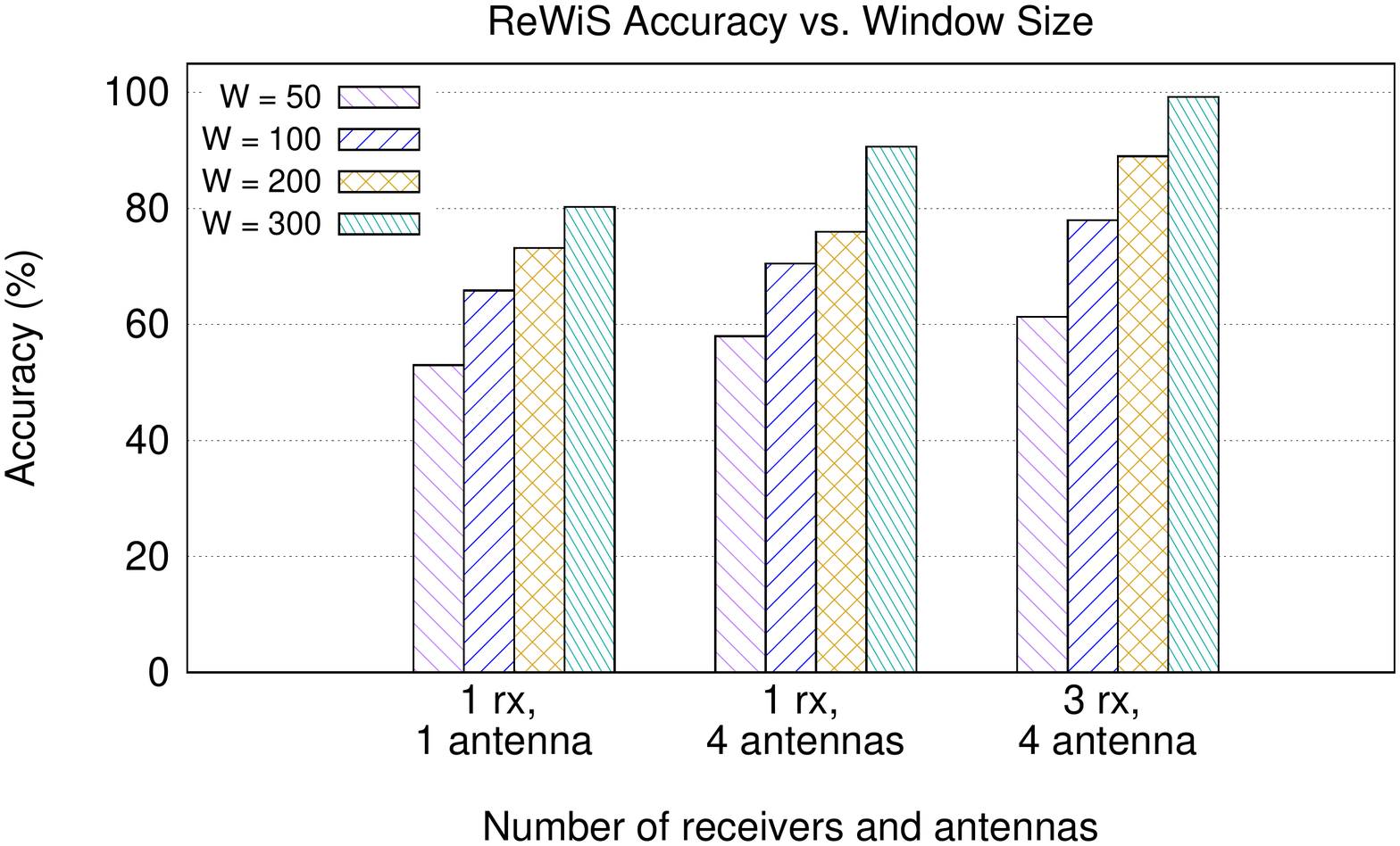}
    \caption{Impact of diversity and window size.\vspace{-0.3cm}}
    \label{fig:window_size}
\end{figure}

Figure \ref{fig:window_size} compares the performance of the algorithm with different window sizes at 80 MHz channel. It can be seen that with a very small window size i.e., $W = 50$ the accuracy of the predictions is very low and it does not improve much by increasing the diversity. 
Specifically, Figure \ref{fig:window_size} shows that (i) by increasing the window size $W$, the number of receivers, and the number of antennas per receiver, the performance improves by up to 35\%, 12\% and 10\%, respectively.
However, with a higher number of antennas, the window size can be decreased and still achieve acceptable accuracy. It is worth noting that reducing the window size reduces pre-processing time and complexity. 

\section{Conclusions and Future Work}\label{sec:conclusion}

In this paper, we have proposed \FW, a novel framework for robust and environment-independent Wi-Fi sensing. \FW leverages multi-antenna multi-receiver diversity to improve the overall robustness, and leverages a customized version of \gls{fsl} to eliminate the need for application-specific feature extraction. \FW has been prototyped using off-the-shelf Wi-Fi equipment, and its performance has been showcased by considering a human activity recognition. We have performed an extensive data collection campaign, evaluated the impact of each diversity component on the performance, and compared \FW with a CNN-based approach. Experimental results have shown that \FW improves the performance by about 40\% with respect to existing single-antenna low-resolution approaches, and increases accuracy by 35\% with respect to a CNN when tested in different environments. As part of the novel contributions, we will release our 60 GB dataset and the entire code repository to the community. We believe that \FW improves the state of the art in Wi-Fi sensing, by demonstrating superior generalization and robustness capabilities with respect to existing work. As part of future work, we are planning to test \FW in the presence of more complex classification tasks, for example, multiple human subjects or additional activities.

\section{Acknowledgment and Disclaimer}

This work is funded in part by the AFRL Visiting Faculty Research Program (VFRP) contract number FA8750-20-3-1003 and National Science Foundation (NSF) grants CNS-2134973 and CNS-2120447.  Any opinions, findings and conclusions or recommendations expressed in this material are those of the authors and do not necessarily reflect the views of the U.S. Government.

\footnotesize
\bibliographystyle{ieeetr}
\bibliography{bib}

\end{document}